\documentclass{pasa}
\usepackage{graphicx}
\usepackage{color}
\usepackage{multirow}
\usepackage{aas_macros}
\usepackage{hyperref}
\hypersetup{colorlinks,citecolor=blue,linkcolor=blue,urlcolor=blue}
\usepackage{pdflscape}
\usepackage{multirow}
\usepackage{ulem}

\def\msun{{\rm M}_{\odot}}

\newcommand{\ijmpa}{Int. J. Mod. Phys. A}

\title[Triple systems and early $r$-process nucleosynthesis]{$r$-process nucleosynthesis in the early Universe through fast mergers of compact binaries in triple systems}

\author[M. Bonetti et al.]{Matteo Bonetti$^{1,2}$\thanks{matteo.bonetti1990@gmail.com}, Albino Perego$^{3,4,5,6,7}$\thanks{albino.perego@mib.infn.it}, Pedro R. Capelo$^8$, Massimo Dotti$^{2,4}$ \& M. Coleman Miller$^{9}$%
\affil{$^1$DiSAT, Universit\`a degli Studi dell'Insubria, Via Valleggio 11, IT-22100 Como, Italy}%
\affil{$^2$INFN, Sezione di Milano-Bicocca, Piazza della Scienza 3, IT-20126 Milano, Italy}%
\affil{$^3$INFN, Sezione di Milano-Bicocca, gruppo collegato di Parma, Parco Area delle Scienze 7/A, IT-43124 Parma, Italy}%
\affil{$^4$Dipartimento di Fisica G. Occhialini, Universit\`a degli Studi di Milano-Bicocca, Piazza della Scienza 3, IT-20126 Milano, Italy}%
\affil{$^5$Dipartimento di Scienze Matematiche Fisiche ed Informatiche, Universit\`a di Parma, Parco Area delle Scienze 7/A, IT-43124 Parma, Italy}%
\affil{$^6$Institut f\"ur Kernphysik, Technische Universit\"at Darmstadt, Schlossgartenstra{\ss}e 2, DE-64289 Darmstadt, Germany}%
\affil{$^7$GSI Helmholtzzentrum f\"ur Schwerionenforschung GmbH, Planckstra{\ss}e 1, DE-64291 Darmstadt, Germany}%
\affil{$^8$Center for Theoretical Astrophysics and Cosmology, Institute for Computational Science, University of Zurich, Winterthurerstrasse 190, CH-8057 Z\"urich, Switzerland}%
\affil{$^9$Department of Astronomy and Joint Space-Science Institute, University of Maryland, College Park, MD 20742-2421, USA}%
}

\jid{PASA}
\doi{10.1017/pas.\the\year.xxx}
\jyear{\the\year}

\begin{document}

\begin{frontmatter}
\maketitle
\begin{abstract}
Surface abundance observations of halo stars hint at the occurrence
of $r$-process nucleosynthesis at low metallicity ($\rm{[Fe/H]< -3}$), 
possibly within the first $10^8$~yr after the formation of the first stars. 
Possible loci of early-Universe $r$-process nucleosynthesis are
the ejecta of either black hole--neutron star or neutron star--neutron star
binary mergers. Here we study the effect of the
inclination--eccentricity oscillations raised by a tertiary (e.g. a star) on the coalescence time-scale of the inner
compact object binaries. Our results are highly sensitive to the
assumed initial distribution of the inner binary semi-major
axes. Distributions with mostly wide compact object binaries are most
affected by the third object, resulting in a strong
increase (by more than a factor of 2) in the fraction of fast coalescences. If instead
the distribution preferentially populates very close compact binaries,
general relativistic precession prevents the third body from increasing the
inner binary eccentricity to very high values. In this last case, the
fraction of coalescing binaries is increased much less by tertiaries,
but the fraction of binaries that would coalesce within $10^8$~yr even
without a third object is already high. Our results provide additional
support to the compact-binary merger scenario for $r$-process
nucleosynthesis.
\end{abstract}
\end{frontmatter}

%%%%%%%%%%%%%%%%%%%%%%%%%%%%%%%%%%%%%%%%%%%%%%%%%%%%%%%%%%%%%%%%%%%%%%
%%%%%%%%%%%%%%%%%%%%%%%%%%%%%%%%%%%%%%%%%%%%%%%%%%%%%%%%%%%%%%%%%%%%%%
\section{Introduction}
\label{sec:intro}

The two main processes responsible for the production of 
the elements beyond iron group nuclei in the Universe are the rapid and slow neutron 
capture processes ($r$-process and $s$-process).
The $s$-process occurs in low- to intermediate-mass stars ($\lesssim 8~\msun$)
during their asymptotic giant branch phase \citep[e.g.][]{Arlandini.etal:1999,Kaeppeler.etal:2011,Karakas.Lattanzio:2014}. 
The duration of the main sequence phase for the stars responsible for the main s-process (1.3--3~$\msun$) 
sets the expected delay ($\gtrsim 0.6~{\rm Gyr}$) for the occurrence of
$s$-process nucleosynthesis in the early Universe \citep[e.g.][]{Sneden2008}.
The site(s) for the $r$-process nucleosynthesis is (are) still debated, as well as the delay
between the formation of the first stars and its first occurrence \citep[see][for a recent 
review]{Thielemann.etal:2017}.

Observations of the surface abundances of old, metal poor stars in the
galactic halo and in nearby dwarf galaxies hint at the occurrence of 
$r$-process nucleosynthesis in the very early stages of cosmological 
evolution \citep{Sneden2003,Honda2006,Sneden2008,Roeder.etal:2014,Ji2016}. The $r$-process occurs 
when the neutron and photon capture rates are higher than the $\beta$-decay 
rate of the unstable capturing nuclei. Therefore,
$r$-process nucleosynthesis requires special conditions to occur, namely a high
neutron-to-seed ratio at Nuclear Statistical Equilibrium
freeze-out \citep[e.g.][]{Hoffman1997a}. These conditions are realized for: 
(i) high neutron densities, 
(ii) expansion time-scales shorter than the neutron lifetime (i.e. explosive environments), 
(iii) neutron-to-proton ratios larger than unity, and
(iv) preferentially high-entropy conditions\footnote{
	If $n_n$ and $n_p$ are the neutron and proton densities, respectively, 
	then for $ n_n/(n_n + n_p) \lesssim 0.25$ $r$-process nucleosynthesis is also effective in 
	synthesizing elements up to the third $r$-process peak for cold, 
	low-entropy matter, i.e $s \lesssim 20 \, k_{\rm B}/{\rm baryon} $, where $k_{\rm B}$ is the Boltzmann constant
\citep[see, e.g.][]{Martin2017}.}.

The large scatter in the observed Europium abundance in old metal poor ($[{\rm Fe/H} ] < -3 $)
stars indicates that $r$-process elements must be synthesized in
rare and isolated events that inject a significant amount of heavy
elements into a relatively small amount of gas.
Such gas must undergo star formation before complete elemental mixing 
has occurred over the entire galaxy. The rare high-yield scenario is also supported by
the comparison of plutonium and iron abundances in deep-sea sediments \citep{Hotokezaka2015}.
Inhomogeneous galactic chemical evolution models indicate that, in order to explain
the distribution of europium abundances at low metallicity, the delay
between the first core collapse supernova (CCSN) explosions and the
production of $r$-process elements cannot exceed $\sim 10^8$~yr \citep{Argast2004,Cescutti2015,Wehmeyer2015},
if efficient galactic mixing is assumed \citep[see however,][for different conclusions based on different modelling and assumptions about the mixing of the ejecta with the 
interstellar medium]{vandeVoort.etal:2015,Shen2015,Hirai.etal:2015}.

According to recent models, the necessary conditions for the occurrence
of $r$-process nucleosynthesis are not reached in standard CCSNe 
\citep[e.g.][and references therein]{Arcones2013}, 
whereas magnetically driven CCSNe could potentially enrich the interstellar 
medium with neutron-rich ejecta. These SNe are
expected to be rare and to inject $10^{-4}$--$10^{-3}$ M$_\odot$ of
$r$-process material per SN \citep{Fujimoto2008,Winteler2012,Nishimura2015}. The presence of rapidly rotating
stellar cores, which are needed for these explosions, is more likely realized at
lower metallicity \citep{Woosley2006a} 
and suggests a possible connection with hypernovae 
and long gamma-ray bursts. Unfortunately, details of the magnetically driven
CCSN explosion mechanism and even the existence of such explosions are still debated \citep[e.g.][]{Moesta2014}.

Another possible site for $r$-process nucleosynthesis in the Universe
are compact-binary mergers (CBMs), with at least one binary component
being a neutron star (NS) \citep{Lattimer1974,Symbalisty1982,Eichler1989,Freiburghaus1999a}. 
This long-standing conjecture has been recently
confirmed by the combined electromagnetic and gravitational wave (GW) detection
from a likely binary NS merger \citep[e.g.][]{Abbott2017,Abbott2017f,Pian2017,Tanvir2017,Coulter2017a,Nicholl2017,Chornock2017}.
The electromagnetic signal is compatible with a kilonova emission, which is thought to be powered 
by the radioactive decay of the freshly synthesized $r$-process elements \citep[e.g.][for recent reviews]{Rosswog2015,Fernandez2016,Metzger2017}.
CBMs can eject $10^{-4}$--$10^{-2}$ M$_\odot$ per merger event 
in the form of dynamical, viscous, 
neutrino-driven or magnetically driven ejecta, although the
precise amount of ejecta depends on the intrinsic properties of the
merging binary, as well as on the still unknown properties of the
nuclear equation of state above nuclear saturation density
\citep[see, e.g.][for some recent discussions]{Surman2008,Korobkin2012,Hotokezaka2013,Fernandez2013,Bauswein2013a,Wanajo2014,Perego2014b,Foucart2015,Martin2015,Just2015,Wu2016,Radice2016,Roberts2017,Bovard2017}.

CBMs are driven by the emission of GWs.
However, the corresponding merger time-scale 
in an isolated binary depends strongly
 on the initial orbital parameters of the compact
binary. Fast (i.e. within $10^8$ yr) binary mergers require small
orbital separations and/or high eccentricities \citep{Peters_1964}. For this
reason the possibility for CBMs to be a viable site for the $r$-process
nucleosynthesis in the early Universe is still disputed.

The strong constraints on the initial semi-major axis and eccentricity for there to be fast coalescence are relaxed if the binary interacts with other objects. The occurrence of such triple or
multiple systems is not negligible: a significant fraction
of massive stars ($M \gtrsim 8 {\rm M_{\odot}}$, whose SN explosion produces
a NS or black hole $-$ BH $-$ remnant) are bound in multiple 
systems \citep[e.g.][]{Duchne2013}.
In the presence of a third object, the stellar system can undergo
Kozai--Lidov (KL) oscillations \citep{Kozai1962,Lidov1962}, in which the eccentricity and
inclination of the inner binary oscillate with periods significantly
longer than the inner orbital period.
Depending on the triplet configuration, the inner binary can increase
 its eccentricity significantly, which then decreases the time to
coalescence due to GW emission.

The effects of the KL mechanism have been invoked in many different astrophysical contexts including: planetary dynamics \citep{Holman1997,Ford_et_al_2000,Katz2011,Naoz2012,Naoz2013}, interactions of stellar size objects in globular clusters \citep{Antonini2016, Antognini2016} and around massive BHs \citep{Antonini2012,VanLandingham2016}, and triple massive BH systems \citep{Miller2002,Blaes_et_al_2002,Iwasawa2006,Hoffman2007,Kulkarni2012,Bonetti2016}.

In a previous work similar in spirit, \citet{Thompson2011} showed that the rate of CBMs can be significantly
enhanced by the KL mechanism within a Hubble time.
In this paper, we explore under which conditions the KL mechanism can affect the dynamics of 
a triplet hosting an inner compact binary, such that the coalescence time-scale becomes shorter than 100~Myr.

The paper is structured as follows. In Section~\ref{sec:estimates}, we introduce the parameters 
involved in our calculations, perform basic estimates, and present the most relevant time-scales. 
We present the equations that describe the triplet evolution in the secular approximation  
in Section~\ref{sec:secular evolution}. Section~\ref{sec:inner binary evolution} is devoted to the 
analysis of the evolution of the inner compact binary in a few selected cases, 
whereas in Section~\ref{sec:distributions}, the effect of the KL mechanism on compact binary populations 
is explored. Finally, we discuss our results and conclude in Section~\ref{sec:conclusions}. 
In Appendix~\ref{app:parameter_exploration}, we summarize and discuss the results of our extensive parameter space exploration.

%%%%%%%%%%%%%%%%%%%%%%%%%%%%%%%%%%%%%%%%%%%%%%%%%%%%%%%%%%%%%%%%%%%%%%
\section{Preliminary estimates and time-scales}\label{sec:estimates}

%%%%%%%%%%%%%%%%%%%%%%%%%%%%%%%%%%%%%%%
\begin{figure*}
\begin{minipage}[c]{0.47\textwidth}
   \centering
   \includegraphics[scale=0.28]{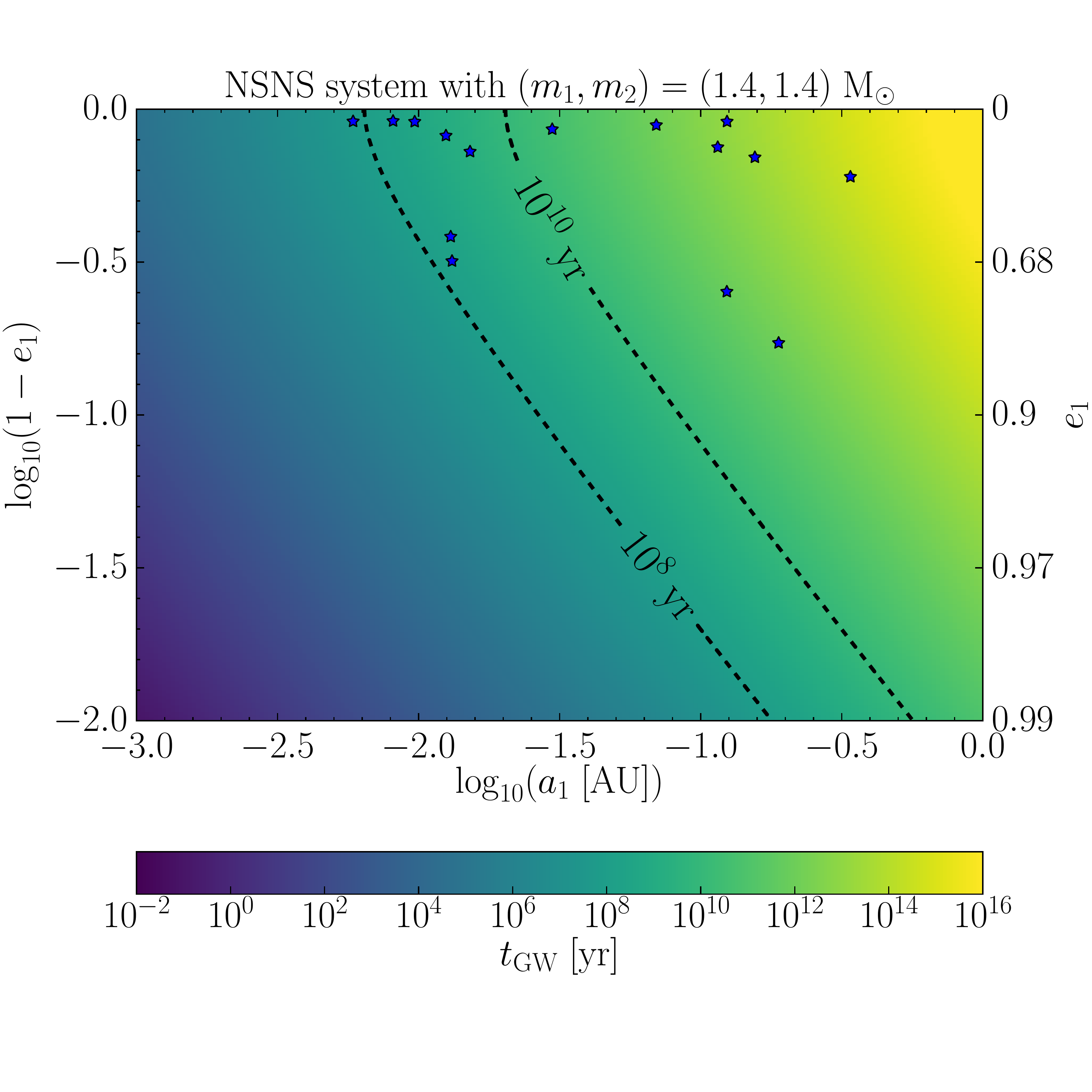}
 \end{minipage}
 \ \hspace{0.5mm} \
 \begin{minipage}[c]{0.47\textwidth}
  \centering
   \includegraphics[scale=0.28]{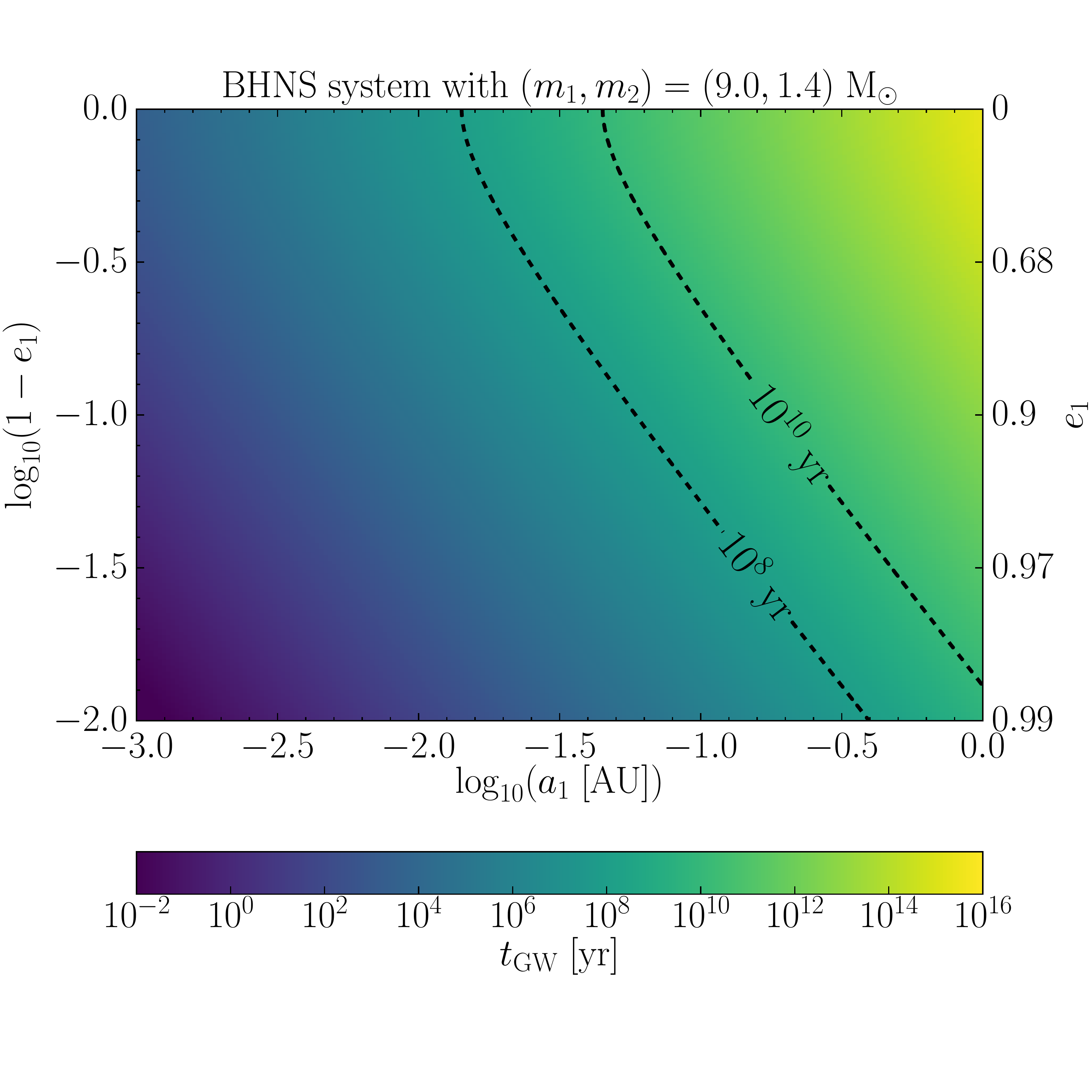}
 \end{minipage}
 \caption{Merger time-scale of an isolated binary due to emission of GWs, as a function of the initial semi-major axis $a_1$ and eccentricity $e_1$.
 {\it Left panel}: NS binary with masses $m_1=m_2=1.4 \ \rm M_{\odot}$. Blue stars refer to the measured or estimated orbital properties of observed NSNS systems
 (see Table~\ref{tab: DNS properties} for more details).  
 {\it Right panel}: BHNS binary with masses $m_1= 9\rm \ M_{\odot}$ and $m_2=1.4 \rm \ M_{\odot}$. 
 Dashed lines mark the values of semi-major axis and eccentricity for which the coalescence takes place within $10^8$ and $10^{10}$~yr.}
 \label{fig:merge_time binary}
\end{figure*}
%%%%%%%%%%%%%%%%%%%%%%%%%%%%%%%%%%%%%%%%

%%%%%%%%%%%%%%%%%%%%%%%%%%%%%%%%%%%%%%%%
\begin{table*}
\centering
\caption{Properties of the observed NSNS systems \citep[adapted from][]{Tauris_et_al_2017}. Pulsar name indicates the name of the radio pulsar(s) in the system. Quantities in brackets are assumed. In particular, if $m_2$ is not measured, but $m_1+m_2$ is, $m_2 = 1.28 \ \msun$ is assumed (central value of the measured secondary mass distribution; 
for B1930-1852, $m_2=1.29 \msun$ to be compatible with observational limits). If also $m_1+m_2$ is not measured, $m_1+m_2=2.725 \ \msun$ is assumed (central value of the measured total mass distribution). The semi-major axis $a_1$ is computed assuming a Keplerian orbit. In the location column, GF and GC stand for galactic field and globular cluster, respectively.}
 \begin{tabular}{ c | c | c | c | c | c | c | c | c }
 \hline
  \multirow{2}{*}{Pulsar name} & $T_{\rm orb}$ & $e_1$    & $m_1$        & $m_2$         & $m_1+m_2$& $a_1$       & \multirow{2}{*}{Location} & $t_{\rm GW}$  \\
             & $[{\rm days}]$& $[-]$  & $[\msun]$    & $[\msun]$     & $[\msun]$    & $[10^{-2} {\rm AU}]$ &  & $[{\rm yr}]$ \\ \hline \hline
  & & & & & & & & \\
  J0453+1559  & 4.072         & 0.113  & 1.559        & 1.774         & 2.734        & 6.959     & GF    & 1.44$\times 10^{12}$    \\
  J0737-3039  & 0.102         & 0.088  & 1.338        & 1.249         & 2.587        & 0.586     & GF    & 8.51$\times 10^{7}$    \\
  J1518+4904  & 8.634         & 0.249  & (1.428)      & (1.28)        & 2.718        & 11.49     & GF    & (8.67$\times 10^{12}$)    \\
  B1534+12    & 0.421         & 0.274  & 1.346        & 1.333         & 2.678        & 1.522     & GF    & 2.71$\times 10^{9}$      \\
  J1753-2240  & 13.638        & 0.304  & (1.445)      & (1.28)        & (2.725)      & (15.562)  & GF    & (2.63$\times 10^{13}$)   \\
  J1755-2550  & 9.696         & 0.089  & (1.445)      & (1.28)        & (2.725)      & (12.40)   & GF    & (1.46$\times 10^{13}$) \\
  J1756-2251  & 0.320         & 0.181  & 1.341        & 1.230         & 2.570        & 1.250     & GF    & 1.64$\times 10^{9}$   \\
  J1811-1736  & 18.779        & 0.828  & <1.64 (1.29) & > 0.93 (1.28) & 2.57         & 18.89     & GF    & (1.78$\times 10^{12}$)     \\
  J1829+2456  & 1.176         & 0.139  & <1.38 (1.31) & > 1.22 (1.28) & 2.59         & 2.976     & GF    & (5.40$\times 10^{10}$)     \\
  J1906+0746  & 0.166         & 0.085  & 1.291        & 1.322         & 2.613        & 0.812     & GF    & 3.05$\times 10^{8}$  \\
  J1913+1102  & 0.206         & 0.090  & <1.84 (1.60) & > 1.04 (1.28) & 2.88         & 0.969     & GF    & (4.65$\times 10^{8}$)           \\
  B1913+16    & 0.323         & 0.617  & 1.440        & 1.389         & 2.828        & 1.299     & GF    & 2.98$\times 10^{8}$     \\
  B1930-1852  & 45.060        & 0.399  & >1.30 (1.30) & < 1.32 (1.29) & 2.59         & 33.94     & GF    & (5.26$\times 10^{14}$)          \\
  B1807-2500B & 9.957         & 0.747  & 1.366        & 1.206         & 2.572        & 12.38     & GC       & 1.03$\times 10^{12}$   \\
  B2127+11C   & 0.335         & 0.681  & 1.358        & 1.354         & 2.713        & 1.314     & GC       & 2.14$\times 10^{8}$    \\
  \hline  
  \hline
 \end{tabular}
 \label{tab: DNS properties}
\end{table*}
%%%%%%%%%%%%%%%%%%%%%%%%%%%%%%%%%%%%%%%%

For an isolated binary system, the merger time-scale is given by the gravitational radiation time, $t_{\rm GW}$, obtained by integrating the coupled evolution of
the semi-major axis and of the inner eccentricity \citep[see, e.g.][]{Peters_1964}. If $m_1$ and $m_2$ (with $q \equiv m_2/m_1 \leq 1$) are the masses of the two bodies orbiting each other and emitting GWs,

%%%%%%%%%%%%%%%%%%%%%%%%%%%%%%%%%%%%%%%%
\begin{align} \label{eq:t_gw}
	t_{\rm GW} = 3.2452\times & 10^8 \ {\rm yr} \ \left( \dfrac{a_1}{0.01 {\rm AU}} \right)^4\nonumber\\
	& \left( \dfrac{\mu_{\rm CB}}{ {\rm M_{\odot}}} \right)^{-1} \left( \dfrac{m_1+m_2}{5 {\rm M_{\odot}}} \right)^{-2} f(e_1),
\end{align}
%%%%%%%%%%%%%%%%%%%%%%%%%%%%%%%%%%%%%%%%
where $a_1$ is the semi-major axis of the initial orbit, $e_1$ its eccentricity, $\mu_{\rm CB} = m_1 m_2 / (m_1+m_2)$ the reduced mass of the inner compact binary, and $f(e_1)$ is a sensitive function of the initial eccentricity:

%%%%%%%%%%%%%%%%%%%%%%%%%%%%%%%%%%%%%%%%
\begin{align}
	f(e_1) &= \biggl[ \dfrac{1-e_1^2}{e_1^{12/19}} \biggl( 1+\dfrac{121}{304}e_1^2 \biggr)^{-870/2299} \biggr]^4\nonumber\\
	& \times \int_0^{e_1} {\rm d}\bar{e} \dfrac{\bar{e}^{29/19}}{(1-\bar{e}^2)^{3/2}} \biggl(1+\dfrac{121}{304}\bar{e}^2\biggr)^{1181/2299}.
\end{align}
%%%%%%%%%%%%%%%%%%%%%%%%%%%%%%%%%%%%%%%%
Following \cite{Peters_1964}, expansions of $f(e)$ can be computed for $e_1 \rightarrow 0$, $f(e_1) \approx (19/48) \left[ (1-e_1^2) \left( 1 + 121 e_1^2/304  \right) \right]^4 $,
and for $e_1 \rightarrow 1$, $f(e_1) \approx (304/425) \left( 1-e_1^2 \right)^{7/2}$. We find that a good approximation over the whole range of $e_1$ is provided by 
$f(e_1) = (1-e_1^2)^{(8-e_1)/2} g(e_1) $, where $g(e_1)$ is a monotonically increasing function varying between $g(0)=19/48 $ and $g(1)= 304/425$.\footnote{A hyperbolic fit
$g(x)=0.38+1/[49.3(-x+1.08)]$ provides an expression accurate to within 1\% between $0 < x < 0.99$.}

In Figure~\ref{fig:merge_time binary}, we present the GW time-scale (equation~\ref{eq:t_gw})
%\citep[numerically computed, according to the expression reported in][]{Peters_1964} 
as a function of $a_1$ and $e_1$ for a typical binary NS (NSNS) system characterized by $m_1 = m_2 = 1.4 \ \msun$ (left panel) and for a black hole--neutron star (BHNS) binary system with $m_1 = 9 \ \msun$ and $m_2 = 1.4 \ \msun$ (right panel).
Clearly, $t_{\rm GW}$ depends strongly on the orbital parameters. In the case of binary NS systems, we report also the orbital properties of the observed NSNS systems \citep[see][and also Table~\ref{tab: DNS properties}]{Tauris_et_al_2017}. Due to the narrow distributions of NS masses in NSNS systems, the calculation of $t_{\rm GW}$ for our reference case ($m_1 = m_2 = 1.4~\msun$) provides an accurate enough estimate also for the merger time-scales of the observed sample of NSNS binaries. Amongst the observed systems, $t_{\rm GW}$ is $<10^8$~yr in only one case, whereas many systems will not coalesce within a Hubble time. A fast merger time-scale (of the order of or below $10^8$~yr) requires a small orbit, $ a_1 \lesssim 0.01~{\rm AU}$, or at larger separations ($a_1 \sim 0.2~{\rm AU}$) a very high eccentricity, $e_1 \gtrsim 0.99$. Due to the larger mass of the BH, the GW time-scale is significantly smaller for BHNS systems at a fixed separation. However, fast mergers still require small orbits or high eccentricities. The lack of observations for such systems prevents a direct comparison with orbital configurations realized in nature.

\begin{figure*}
	\begin{minipage}[c]{0.47\textwidth}
		\centering
		\includegraphics[scale=0.24]{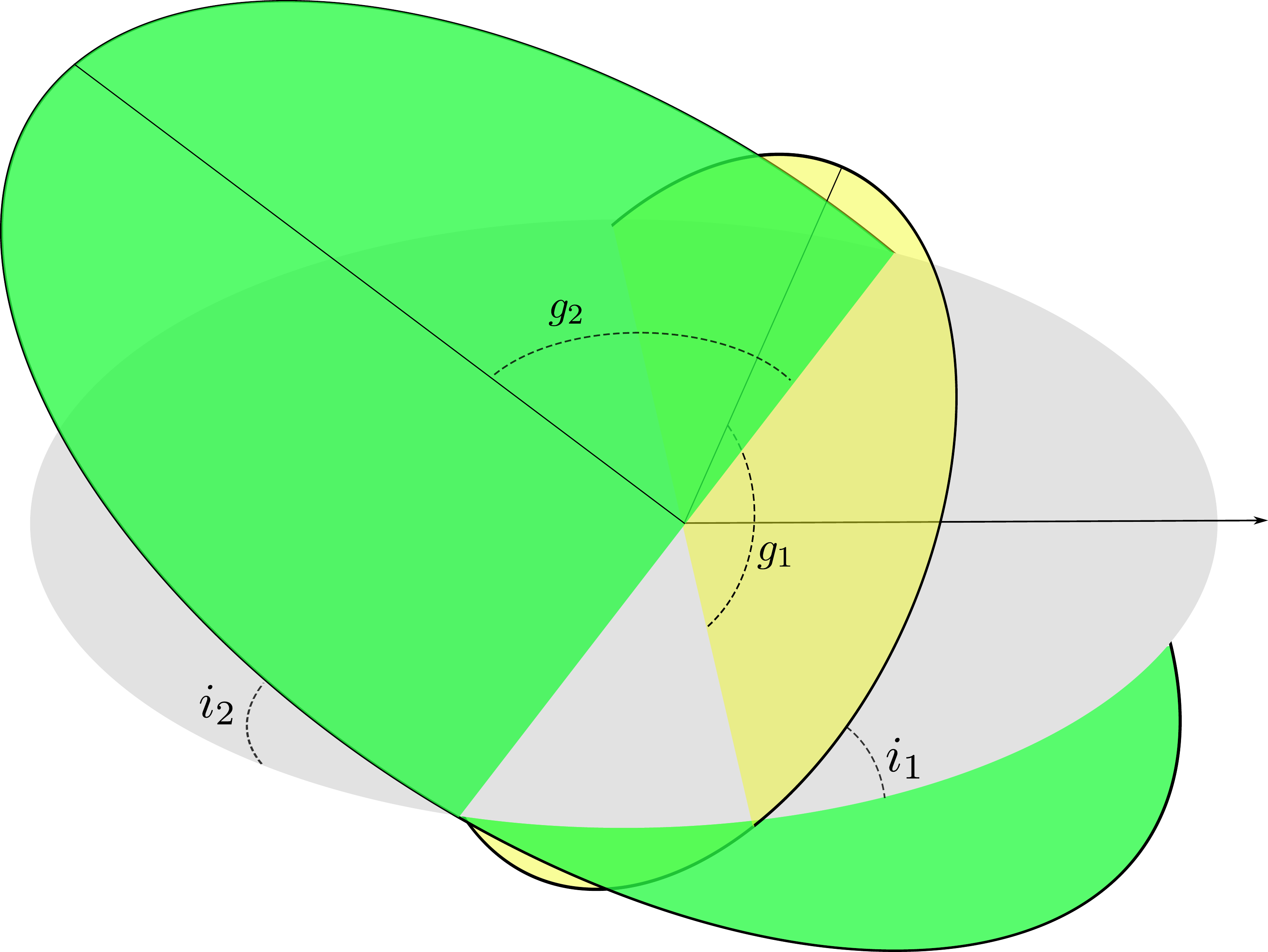}
	\end{minipage}
	\ \hspace{0.5mm} \
	\begin{minipage}[c]{0.47\textwidth}
		\centering
		\includegraphics[scale=0.45]{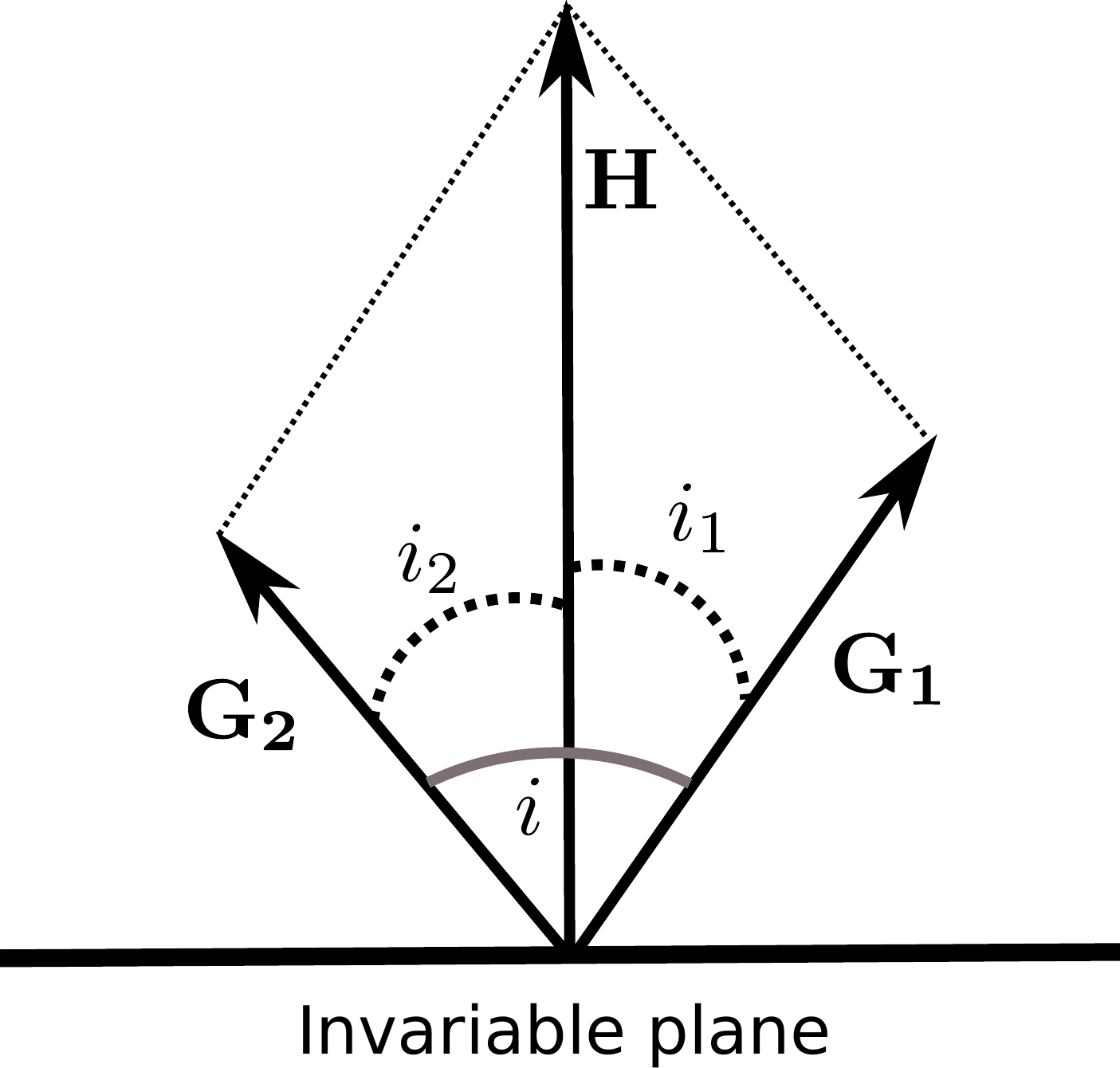}
	\end{minipage}
	\caption{Schematic description of the configuration of hierarchical triplets. {\it Left panel}: configuration in the 3D space. {\it Right panel}: configuration of the angular momenta. Note that the definition of the relative inclination $i \equiv i_1+i_2$ results rather natural.}
	\label{fig:angles}
\end{figure*}

If the compact binary is part of a gravitationally bound triple system, its properties are fully specified once the positions, velocities, and masses of the three bodies are known at one instant in time. We restrict our study to the case where the triplet is hierarchical and its evolution is well described by a secular approach. Under these hypotheses, the description of the triplet is simplified because it can be treated as consisting of two distinct, but coupled, binary systems:

(i) an inner binary, which in our case is always represented by a compact binary and is characterized by the following minimal 
set of six parameters:

%%%%%%%%%%%%%%%%%%%%%%%%%%%%%%%%%%%%%%%%
\begin{itemize}

\item $a_1$, the inner semi-major axis, such that $ 5 \times 10^{-3}~{\rm AU} < a_1 < 0.3~{\rm AU} $, which is compatible with the observed NSNS semi-major axes. We also include the possibility that $a_1$ is smaller than what is currently observed, because a population of tight compact binaries could be difficult to observe, due to the short $t_{\rm GW}$;

\item $e_1$, the inner eccentricity, such that $ 0 < e_1 < 1 $;

\item the primary and secondary masses, $m_1$ and $m_2$. For NSs, we consider $ 1.0~\msun < m_{\rm NS} < 2.4~\msun$, which is $\sim20\%$ wider than the maximum and minimum observed NS masses; for BHs, we choose $5~\msun < m_{\rm BH} < 30~\msun$, which is within the highly uncertain range of stellar BH masses observed in binaries;

\item the inner argument of the pericentre, $g_1$, which locates the angular position of the pericentre in the orbital plane and is between 0 and $2\pi$ radians (see left panel of Figure~\ref{fig:angles});

\item the inner inclination angle, $i_1$, which is the angle between the positive $z$ direction and the orbital angular momentum of the inner binary, $\mathbf{G}_1$, i.e. $\cos{i_1} = \mathbf{G}_1\cdot \mathbf{\hat{z}} / G_1 $, where $ \mathbf{\hat{z}} $ is the unitary positive vector along $z$ (where we define $z$ to be along the direction of the total angular momentum, ${\bf H}={\bf G}_1+{\bf G}_2=H \mathbf{\hat{z}}$). Thus, in general $0 \leq i_1 \leq \pi $ and $i_1 < \pi/2$ represents counter-clockwise motion (see right panel of Figure~\ref{fig:angles}).

\end{itemize}
%%%%%%%%%%%%%%%%%%%%%%%%%%%%%%%%%%%%%%%%

%%%%%%%%%%%%%%%%%%%%%%%%%%%%%%%%%%%%%%%
%%%%%%%%%%%%%%%%%%%%%%%%%%%%%%%%%%%%%%%%

(ii) an outer binary system, in which the inner binary is treated as a point of mass $m_1+m_2$, located in its centre of mass, and the second component is a main sequence star of mass $m_3$. The outer binary is characterized by a set of five parameters, similar to that of the inner binary:

%%%%%%%%%%%%%%%%%%%%%%%%%%%%%%%%%%%%%%%%
\begin{itemize}

\item $a_2$, the outer semi-major axis, such that $ 3 \times 10^{-2}~{\rm AU} < a_1 < 10~{\rm AU}$. Observed external semi-major axes of hierarchical triple stellar systems span a wide range of values, going from a fraction of AU up to thousands of AU. We impose an upper limit of 10~AU to ensure a significant coupling between the inner and the outer binary;

\item $e_2$, the outer eccentricity, such that $ 0 < e_2 < 1 $;

\item the tertiary mass, $m_3$, with $ 3~\msun < m_3 < 15~\msun$. The lower limit on $m_3$ is required to have an adequate gravitational influence on the dynamics of the inner binary, whose total mass is always above $2~\msun$. Our choice is also supported by the fact that stars in the early Universe are metal-poor and therefore more massive \citep[e.g.][]{Bromm.etal:2002}. Moreover, hierarchical triplets with light tertiary masses are easier to unbind by external perturbations.
The upper limit is related to the stability of the triplet itself. Indeed, the presence of a main sequence star requires consideration of the stellar main-sequence lifetime:
 
\begin{equation}
t_{\rm MS} \sim 10^{10} \, {\rm yr} \left( \frac{m_3}{\msun} \right)^{-5/2} \ .
\label{eq:stellar lifetime}
\end{equation}

For durations greater than $t_{\rm MS}$, the formation of a white dwarf or the explosion of the star as a CCSN can significantly alter the properties of the triplet or even destroy it. Since we are interested in time intervals less than $10^8 \, {\rm yr}$, we use an upper limit for $m_3$ such that $t_{\rm MS}$ equals $10^7 \, {\rm yr}$, i.e. $10\%$ of the maximum allowed time. This corresponds roughly to $16 \ \msun$; we also notice that $t_{\rm MS} \sim 10^{8} \, {\rm yr}$ corresponds to $m_3 \approx 6.3 \ \msun$;

\item the outer argument of the pericentre, $g_2$, which like $g_1$ can vary over $2 \pi$ (see left panel of Figure~\ref{fig:angles});

\item the outer inclination angle, $i_2$, analogous to $i_1$, but for the outer orbit: $\cos{i_2} = \mathbf{G}_2\cdot \mathbf{\hat{z}} / G_2 $, where $\mathbf{G}_2$ is the orbital angular momentum of the outer binary (see right panel of Figure~\ref{fig:angles}).
 
\end{itemize}
%%%%%%%%%%%%%%%%%%%%%%%%%%%%%%%%%%%%%%%%

The only relevant inclination angle is the relative angle between the inner and the outer binaries, $i \equiv i_2 + i_1$. Hence, the hierarchical triplet is characterized by a set of ten independent parameters.

The hierarchical nature of the triplet and the validity of our secular approach constrain the values of the allowed orbital parameters. In particular, we require that  our triplets satisfy the stability criterion reported by \citet{Mardling_Aarseth_2001}:

%%%%%%%%%%%%%%%%%%%%%%%%%%%%%%%%%%%%%%%%
\begin{equation}
 \frac{a_2}{a_1} > 2.8 \, \left( 1 + \frac{m_3}{m_1+m_2} \right)^{2/5} \frac{\left( 1 + e_2  \right)^{2/5}}{\left(  1 - e_2  \right)^{6/5}} \ .
 \label{eq:stability criterium}
\end{equation}
%%%%%%%%%%%%%%%%%%%%%%%%%%%%%%%%%%%%%%%%

This relation was obtained for purely Newtonian coplanar prograde orbits of the inner and outer binaries. Inclined and retrograde orbits are expected to be more stable \citep{Mardling_Aarseth_2001},\footnote{For misaligned orbits, the critical outer semi-major axis for which a triplet remains stable can be reduced by a factor $k = 1-0.3\ i / \pi$ \citep[see][and reference therein]{Mardling_Aarseth_2001}. Note that the minimum allowed $a_2$ is achieved for coplanar retrograde systems.} so equation~(\ref{eq:stability criterium}) provides a conservative stability limit. We assume that triplets for which equation~(\ref{eq:stability criterium}) is not satisfied cannot be treated with the secular approximation and enter the chaotic regime. The precise evolution of such systems requires direct integration of the equations of motion for the three bodies \citep[see, e.g.][and references therein]{Hoffman2007,Antonini2016,Bonetti2016}. In the following, we will assume that in those cases the triplet usually gets disrupted and that the more massive third body probably replaces the lighter NS in the inner binary. Thus, those systems will never host a compact binary merger.

A hierarchical triplet is potentially subject to a large variety of effects that influence its dynamics \citep{Heggie:1975}. Assuming that the triple system is not influenced by dynamical interactions with other external bodies, the most important effects are the general relativistic (GR) precession of the inner periastron and the KL mechanism. The GR precession forces the argument of pericentre of a binary to monotonically increase from 0 to $2\pi$, i.e. the ellipse rotates in the orbital plane and describes rosetta-like orbits, on a time-scale that is given approximately by \citep{Miller2002,Blaes_et_al_2002}

%%%%%%%%%%%%%%%%%%%%%%%%%%%%%%%%%%%%%%%%
\begin{align}
 t_{\rm GR,prec} & \sim 30 \, {\rm yr} \, \left( \frac{m_1+m_2}{5 \, \msun} \right)^{-3/2} \nonumber \\
                    &     \left( \frac{a_1}{0.01 \, {\rm AU}} \right)^{5/2} \left(1-e_1^2 \right) \, .
 \label{eq:GR precession time-scale}
\end{align}
%%%%%%%%%%%%%%%%%%%%%%%%%%%%%%%%%%%%%%%%

If the mutual inclination angle $i$ is large enough, the KL mechanism can induce an oscillation in the inner eccentricity. If we consider the limit\footnote{This condition actually means that the total angular momentum of the system is dominated by the outer binary.} $m_2 \rightarrow 0$ and the first non-vanishing contribution (i.e. the quadrupole term) in the $a_1/a_2$ expansion of the equations of motion, we obtain the classical KL mechanism and $e_1$ oscillates up to a maximum value given by

%%%%%%%%%%%%%%%%%%%%%%%%%%%%%%%%%%%%%%%%
\begin{equation} \label{eq: max KL eccentricity}
 e_{1,{\rm max}} \approx \left( 1 - \frac{5}{3} \cos^2{i} \right)^{1/2} \, 
\end{equation}
%%%%%%%%%%%%%%%%%%%%%%%%%%%%%%%%%%%%%%%%
on a characteristic time-scale

%%%%%%%%%%%%%%%%%%%%%%%%%%%%%%%%%%%%%%%%
\begin{align}
  t_{\rm KL,quad} & \sim  0.4 \, {\rm yr} \, \left( \frac{a_1}{0.01 \, {\rm AU}} \right)^{-3/2} \left( \frac{m_1+m_2}{5 \, \msun} \right)^{1/2} \nonumber \\
      &   \left( \frac{m_3}{10 \, \msun} \right)^{-1} \left( 1-e_2^2 \right)^{3/2} \left( \frac{a_2}{0.1 \, {\rm AU}} \right)^{3} \, .
  \label{eq:KL quadrupole time scale}
\end{align}
%%%%%%%%%%%%%%%%%%%%%%%%%%%%%%%%%%%%%%%%

If $ t_{\rm GR,prec} \lesssim t_{\rm KL,quad} $, the GR precession can erase the KL resonance because it destroys the coherent piling up of the perturbation induced by the third body. Because of the GR precession the maximum eccentricity reached can be much lower \citep{Miller2002}. Using equations~(\ref{eq:GR precession time-scale}) and~(\ref{eq:KL quadrupole time scale}), we obtain a criterion on the orbital parameters for the KL mechanism to be efficient against the GR precession:

%%%%%%%%%%%%%%%%%%%%%%%%%%%%%%%%%%%%%%%%
\begin{align}
  a_2 & < 0.53 \, {\rm AU} \, \left( \frac{a_1}{0.01 \, {\rm AU}} \right)^{4/3} \left( \frac{m_1+m_2}{5 \, \msun} \right)^{-1/3} \nonumber \\
      & \left( \frac{m_3}{m_1+m_2} \right)^{1/3} \left( \frac{1-e_1^2}{1-e_2^2} \right)^{1/2} \, .
  \label{eq:precession criterium}
\end{align}
%%%%%%%%%%%%%%%%%%%%%%%%%%%%%%%%%%%%%%%%

If the KL resonance is not suppressed, the octupole term in the $a_1/a_2$ expansion modulates the $e_1$ oscillation, on a longer time-scale given by

%%%%%%%%%%%%%%%%%%%%%%%%%%%%%%%%%%%%%%%%
\begin{align}
  t_{\rm KL,oct} & \sim 5.3 \, {\rm yr} \, \left( \frac{a_1}{0.01 \, {\rm AU}} \right)^{-5/2} \left( \frac{m_1+m_2}{5 \, \msun} \right)^{3/2} \nonumber \\
      & \left( \frac{m_3}{10 \, \msun} \right)^{-1} \frac{\left( 1-e_2^2 \right)^{5/2}}{e_2} \left( \frac{a_2}{0.1 \, {\rm AU}} \right)^{4} \nonumber \\
      & \left( \frac{\left| m_1 - m_2   \right|}{1 \, \msun} \right)^{-1} \, .
  \label{eq:KL octupole time scale}
\end{align}
%%%%%%%%%%%%%%%%%%%%%%%%%%%%%%%%%%%%%%%%
The effect of the octupole modulation is to increase $e_{1,{\rm max}}$.

%%%%%%%%%%%%%%%%%%%%%%%%%%%%%%%%%%%%%%%%%%%%%%%%%%%%%%%%%%%%%%%%%%%%%%
\section{Secular evolution of isolated hierarchical triplets}\label{sec:secular evolution}

The evolution of the orbital elements of the inner ($a_1$, $e_1$, and $g_1$) and outer ($e_2$ and $g_2$)\footnote{Here we are neglecting the effect of GW emission on the shrinking of the outer binary, hence $a_2$ remains constant throughout the integration.} binaries is obtained under two approximations: $(i)$ the properties of each binary are orbitally averaged, and $(ii)$ the equations of motion are approximated with their expansion up to the second order (octupole term) in $a_1/a_2$.  In detail, we follow \citet{Blaes_et_al_2002} by integrating the following differential equations:

%%%%%%%%%%%%%%%%%%%%%%%%%%%%%
\begin{equation}
\frac{da_1}{dt}=-\frac{64G^3m_1m_2(m_1+m_2)}{5c^5a_1^3(1-e_1^2)^{7/2}}\left(1+\frac{73}{24}e_1^2+\frac{37}{96}e_1^4\right),
\label{eq:da1dt}
\end{equation}
%%%%%%%%%%%%%%%%%%%%%%%%%%%%%

%%%%%%%%%%%%%%%%%%%%%%%%%%%%%
\begin{align}
\frac{dg_1}{dt}&=6C_2\left\{\frac{1}{G_1}[4\cos{i}^2+(5\cos2g_1-1)(1-e_1^2-\cos^2{i})]\right. \nonumber \\
&  \left. +\frac{\cos{i}}{G_2}[2+e_1^2(3-5\cos2g_1)]\right\} +C_3e_2e_1\left(\frac{1}{G_2}+\frac{\cos{i}}{G_1}\right) \nonumber \\
&  \left\{\sin g_1\sin g_2[A+10(3\cos^2{i}-1)(1-e_1^2)]-5\cos{i}~B\cos\phi\right\}\nonumber\\
& -C_3e_2\frac{1-e_1^2}{e_1G_1}\left[10\cos{i}(1-\cos^2{i})(1-3e_1^2)\sin g_1\sin g_2 \right. \nonumber\\
& \left. +\cos\phi(3A-10\cos^2{i}+2)\right] \nonumber\\
& + \frac{3}{c^2a_1(1-e_1^2)}\left[\frac{G(m_1+m_2)}{a_1}\right]^{3/2},
\label{eq:dg1dt}
\end{align}
%%%%%%%%%%%%%%%%%%%%%%%%%%%%%

%%%%%%%%%%%%%%%%%%%%%%%%%%%%%
\begin{align}
\frac{de_1}{dt}&=30C_2\frac{e_1(1-e_1^2)}{G_1}(1-\cos^2{i})\sin2g_1\nonumber\\
& -C_3e_2\frac{1-e_1^2}{G_1}\big[35\cos\phi(1-\cos^2{i})e_1^2\sin2g_1 \nonumber \\
&  -10\cos{i}(1-e_1^2)(1-\cos^2{i})\cos g_1\sin g_2\nonumber\\
& -A(\sin g_1\cos g_2-\cos{i}~\cos g_1\sin g_2)\big]\nonumber\\
& -\frac{304G^3m_1m_2(m_1+m_2)e_1}{15c^5a_1^4(1-e_1^2)^{5/2}}\left(1+\frac{121}{304}e_1^2\right),
\label{eq:de1dt}
\end{align}
%%%%%%%%%%%%%%%%%%%%%%%%%%%%%

%%%%%%%%%%%%%%%%%%%%%%%%%%%%%
\begin{align}
\frac{dg_2}{dt}&=3C_2\biggl\{\frac{2\cos{i}}{G_1}[2+e_1^2(3-5\cos2g_1)]\nonumber\\
&+\frac{1}{G_2}[4+6e_1^2+(5\cos^2{i}-3)(2+3e_1^2-5e_1^2\cos2g_1)]\biggr\}\nonumber\\
& -C_3e_1\sin g_1\sin g_2\bigg\{\frac{4e_2^2+1}{e_2G_2}10\cos{i}~(1-\cos^2{i})(1-e_1^2) \nonumber\\
& -e_2\left(\frac{1}{G_1}+\frac{\cos{i}}{G_2}\right)[A+10(3\cos^2{i}-1)(1-e_1^2)]\bigg\}\nonumber\\
& -C_3e_1\cos\phi\left[5B~\cos{i} e_2\left(\frac{1}{G_1}+\frac{\cos{i}}{G_2}\right)+\frac{4e_2^2+1}{e_2G_2}A\right],
\label{eq:dg2dt}
\end{align}
%%%%%%%%%%%%%%%%%%%%%%%%%%%%%

%%%%%%%%%%%%%%%%%%%%%%%%%%%%%
\begin{align}
\dfrac{de_2}{dt}&=C_3e_1\dfrac{1-e_2^2}{G_2}[10\cos{i}(1-\cos^2{i})(1-e_1^2)\sin g_1\cos g_2\nonumber\\
&+A(\cos g_1\sin g_2-\cos{i}\sin g_1\cos g_2)],
\label{eq:de2dt}
\end{align}
%%%%%%%%%%%%%%%%%%%%%%%%%%%%%

\noindent where $\phi$ is the angle between the periastron directions,

%%%%%%%%%%%%%%%%%%%%%%%%%%%%%
\begin{equation}
\cos{\phi}=-\cos{g_1}\cos{g_2}-\cos{i}\sin{g_1}\sin{g_2},
\end{equation}
%%%%%%%%%%%%%%%%%%%%%%%%%%%%%

\noindent and the cosine of the mutual inclination of the binaries can be expressed as a function of the magnitudes of the  angular momenta of the inner binary ($G_1=m_1m_2\{[Ga_1(1-e_1^2)]/[m_1+m_2]\}^{1/2}$), of the outer binary ($G_2=m_3(m_1+m_2)\{[Ga_2(1-e_2^2)]/[m_1+m_2+m_3]\}^{1/2}$), and of the whole triple system ($H=G_1 \cos{i_1}+G_2 \cos{i_2}$)
% where $i_1$ and $i_2$ are the inclinations of {\bf G$_1$} and {\bf G$_2$} with respect to {\bf H}) 
as follows:

%%%%%%%%%%%%%%%%%%%%%%%%%%%%%
\begin{equation}
\cos i=\frac{H^2-G_1^2-G_2^2}{2G_1G_2}.
\label{eq:eqtheta}
\end{equation}
%%%%%%%%%%%%%%%%%%%%%%%%%%%%%

The closure of the system of differential equations is obtained through the angular momentum evolution equation:

%%%%%%%%%%%%%%%%%%%%%%%%%%%%%
\begin{align}
\dfrac{dH}{dt}&=-\dfrac{32G^3m_1^2m_2^2}{5c^5a_1^3(1-e_1^2)^2}
             \left[\dfrac{G(m_1+m_2)}{a_1}\right]^{1/2}\nonumber\\
             &\left(1+\dfrac{7}{8}e_1^2\right)\dfrac{G_1+G_2\cos{i}}{H}.
\label{eq:dhdt}
\end{align}
%%%%%%%%%%%%%%%%%%%%%%%%%%%%%

In equations~(\ref{eq:dg1dt}--\ref{eq:de2dt}), $A=4+3e_1^2-5(1-\cos^2{i})B/2$ and $B=2+5e_1^2-7e_1^2\cos2g_1$, whereas the quantities $C_2$ and $C_3$ \citep[defined as in][]{Ford_et_al_2000},

%%%%%%%%%%%%%%%%%%%%%%%%%%%%% 
\begin{equation}\label{eq:C2}
C_2={\frac{Gm_1m_2m_3}{16(m_1+m_2)a_2(1-e_2^2)^{3/2}}}\left(\frac{a_1}{a_2}
\right)^2,
\end{equation}
%%%%%%%%%%%%%%%%%%%%%%%%%%%%%

%%%%%%%%%%%%%%%%%%%%%%%%%%%%%
\begin{equation}\label{eq:C3}
C_3=\frac{15Gm_1m_2m_3(m_1-m_2)}{64(m_1+m_2)^2a_2(1-e_2^2)^{5/2}}
\left(\frac{a_1}{a_2}\right)^3,
\end{equation}
%%%%%%%%%%%%%%%%%%%%%%%%%%%%%
%
belong to the quadrupole and octupole terms in the interaction between the two binaries, respectively. All the remaining terms are due to GR effects: the precession of the inner periastron is taken into account in the evolution equation of $g_1$, whereas the back-reaction of GW emission onto the inner binary is included in the evolution equations for $a_1$, $e_1$, and $H$. In particular, if GW emission is neglected, then $dH/dt = 0$, as expected. We stress that such equations are obtained under an approximation that fails for $a_2 \sim a_1$. This does not affect our results, as in this limit the binaries are in the chaotic regime discussed in Section~\ref{sec:estimates}, and are therefore not evolved. Equations~(\ref{eq:da1dt}--\ref{eq:dhdt}) present some interesting symmetries: apart from the trivial invariance for the exchange of the inner binary masses, $m_1'=m_2$ and $m_2'= m_1$, we notice also the invariance under the following transformation of the arguments of periastron: $g_1'=g_1 + \pi$ and $g_2'= g_2 + \pi$.

As a final note, in order to remove the divergence for $e_1\rightarrow 0$ in the octupole term of equation~(\ref{eq:dg1dt}), we solve the system of differential equations above in terms of the auxiliary variables $e_1\cos{g_1}$, $e_1\sin{g_1}$, $e_2\cos{g_2}$, and $e_2\sin{g_2}$, as suggested by \citet{Ford_et_al_2000}.

%%%%%%%%%%%%%%%%%%%%%%%%%%%%%%%%%%%%%%%%%%%%%%%%%%%%%%%%%%%%%%%%%%%%%%
\section{Orbital evolution of inner compact binaries}
\label{sec:inner binary evolution}

%%%%%%%%%%%%%%%%%%%%%%%%%%%%%%%%%%%%%%%%
\begin{figure*}
 \centering
 \includegraphics[scale=0.35]{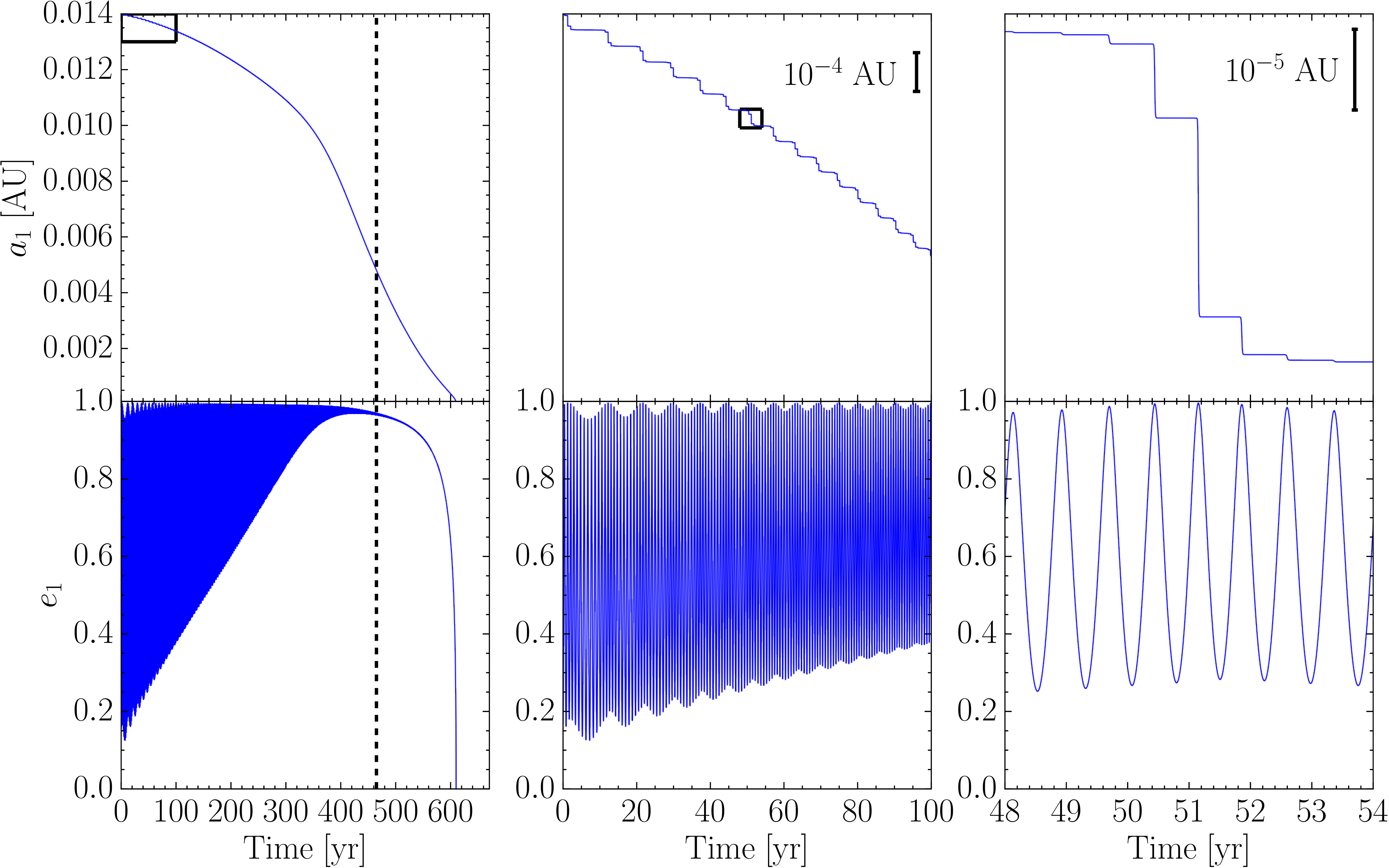}
 \caption{Triplet with a BHNS inner binary. The initial orbital parameters of the inner binary are: $a_1=0.014$~AU, $e_1=0.150$, $m_1 = 9~\msun$, $m_2 = 1.2~\msun$, and $g_1 = 0^\circ$. The outer orbit is characterized by $a_2=0.306$~AU, $e_2=0.6$, $g_2 = 90^\circ $, $i=85^\circ$, and $m_3 = 16~\msun$. {\it Left panels}: full evolution; {\it Central panels}: zoom-in on the octupole time-scale. {\it Right panels}: zoom-in on the quadrupole time-scale. {\it Upper panels}: evolution of the inner binary semi-major axis. {\it Lower panels}: evolution of the inner binary eccentricity. Note the sharp decrease of the semi-major axis when the eccentricity reaches its maximum value. The dashed vertical line corresponds to the point after which the KL mechanism does not significantly influence the evolution and GW emission takes over (see text for more details).}
 \label{fig:BHNS_evolution_example}
\end{figure*}
%%%%%%%%%%%%%%%%%%%%%%%%%%%%%%%%%%%%%%%%
%%%%%%%%%%%%%%%%%%%%%%%%%%%%%%%%%%%%%%%%
\begin{figure*}
 \centering
 \includegraphics[scale=0.35]{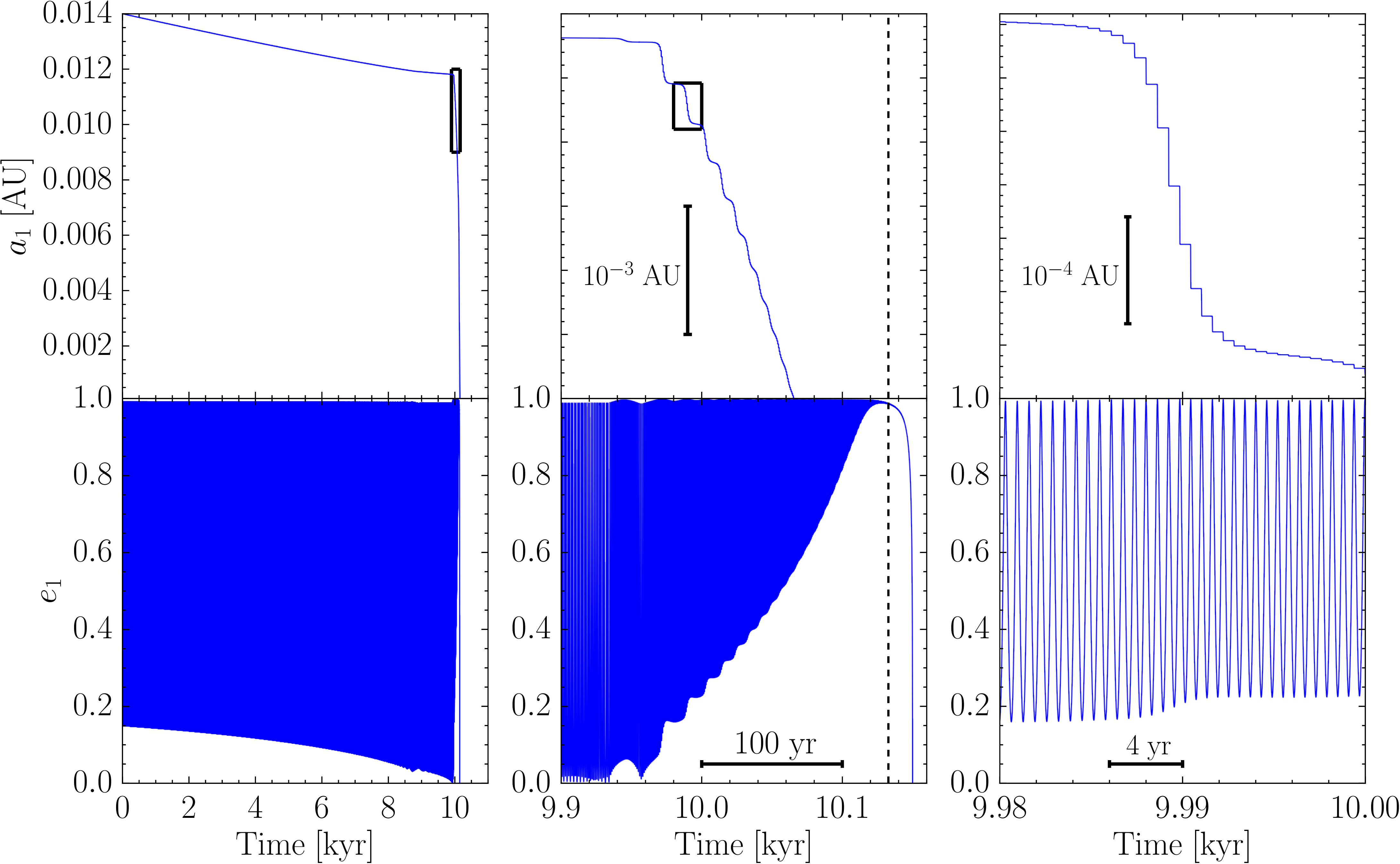}
 \caption{Same as Figure~\ref{fig:BHNS_evolution_example}, except that the inner binary is a NSNS system with masses $(m_1,m_2) = (1.6,1.2) \ \rm M_{\odot}$ and $g_1 = 90^\circ$. Note the change of phenomenology around $t\sim 9.93 \times 10^3$ yr when, because of the octupole term, the argument of pericentre of the inner binary changes from a libration to a circulation regime (see text).}
 \label{fig:NSNS_evolution_example}
\end{figure*}
%%%%%%%%%%%%%%%%%%%%%%%%%%%%%%%%%%%%%%%%

The primary effect of the KL mechanism is the eccentricity growth that the inner binary can experience 
if certain conditions are satisfied. In the standard lore, the trigger conditions are derived with the assumptions that the total angular momentum is dominated by the outer binary and only the quadrupole order of approximation is considered. In this case, if the orbital planes of the inner and outer binary are misaligned, 
with relative inclination in the range $39^\circ \lesssim i \lesssim 141^\circ$
(see equation~\ref{eq: max KL eccentricity}), then secular exchanges of angular momentum between the two 
binaries can excite large oscillations of the relative inclination and of the inner eccentricity. 
When the initial relative inclination is close to $90^\circ$, the process shows its most extreme phenomenology:
during the oscillations, the inner eccentricity can reach values close to unity that can potentially force 
the inner binary to coalesce.\footnote{More precisely, when a relevant fraction of the total angular momentum of the triplet is provided by the inner binary, the condition $e_1 \rightarrow 1$ occurs at relative inclinations greater then $90^\circ$ \citep[see, e.g.][]{Lidov1976,Miller2002}.}

As pointed out in Section~\ref{sec:estimates}, this secular process can be suppressed if the orbit precesses \citep{Holman1997,Ford_et_al_2000,Miller2002,Blaes_et_al_2002}. 
Indeed, the resonance on which the KL mechanism relies strongly depends on the coherent piling up of the 
perturbation exerted by the third body. If the inner binary starts to precess with a time-scale much shorter 
than that of the KL oscillation, then the coherence is destroyed and the process is severely inhibited. 
For compact objects, the most relevant form of precession is the relativistic one. Therefore, in order 
not to overestimate the effect of the KL oscillation, the inclusion of this relativistic effect is crucial. 
In contrast, if the time-scale associated to the KL mechanism is shorter than that of the 
relativistic precession, then the process is only partially perturbed and a triple system can experience 
 eccentricity excitations.

In Figures~\ref{fig:BHNS_evolution_example} and~\ref{fig:NSNS_evolution_example}, we show two representative 
cases that describe the evolution of a BHNS and a NSNS binary, respectively, obtained by integrating
equations~(\ref{eq:da1dt}--\ref{eq:dhdt}). In both cases, the effect of secular evolution 
is clearly visible and drives the compact binary to coalescence within a time much shorter than the coalescence time for
GW emission only.
The upper and lower panels of the two figures show the evolution of the inner semi-major axis and of the inner eccentricity, respectively. 
The left panels describe the whole evolution of the inner compact binary up to coalescence. Note that single 
KL cycles cannot be resolved, as the oscillations proceed on a time-scale much shorter than that of the complete evolution. 
An interesting pattern is clearly visible in the evolution of the eccentricity: as the binary shrinks, 
the minimum inner eccentricity increases. As a consequence, the oscillation range 
of $e_1$ is reduced and the average value of $e_1$ experiences a net increment. 
This is due to the effect of GR corrections, which become stronger as the semi-major axis 
decreases and determine an increase of the minimum value of the relative inclination, which 
in turn increases the minimum eccentricity. This phenomenology persists until the semi-major axis has 
shrunk by nearly one order of magnitude.
At that time, the KL mechanism is not efficient any longer in driving the dynamics of the systems.
Then, the GW emission eventually takes over and quickly drives the binary toward coalescence.  
We mark this point with dashed vertical lines in the left and central panels of 
Figures~\ref{fig:BHNS_evolution_example} and~\ref{fig:NSNS_evolution_example}, respectively.
We computed it as the moment when the residual time to merger and the GW time-scale differ by less than 1\%. Due to the oscillatory behaviour of the eccentricity, for the evaluation 
of the GW time-scale (equation~\ref{eq:t_gw}), we employ orbital elements averaged 
over one quadrupole oscillation, i.e. $t_{\rm GW} = t_{\rm GW}(\langle a \rangle_{\rm KL},\langle e \rangle_{\rm KL})$.

Interesting patterns can be appreciated by zooming into different time-slices of the evolution, 
as represented in the central and right-hand panels. The central panels show a zoom-in on a time length
comparable to the octupole time-scale of the systems, whereas the right-hand panels focus on the quadrupole time-scale. 
When the eccentricity reaches the peak of the quadrupole oscillation with values close to unity (cf. the right-hand panels), 
the semi-major axis decreases sharply as a consequence of an efficient emission of GWs. 
Moreover, the octupole terms (cf. the central panels) clearly modulate the eccentricity growth and push 
its maximum value even further, determining a stronger and sharper extraction of orbital energy 
(cf. right-hand panels, where a sharper decrease of $a_1$ is seen at the peak of the octupole modulation). 
Equations~(\ref{eq:KL quadrupole time scale}) and~(\ref{eq:KL octupole time scale}) provide analytical estimates 
of the quadrupole and octupole time-scales, respectively. The values provided by these expressions for the represented cases are 
$t_{\rm KL, quad} \sim 3.2 \ (2.5)$~yr and $t_{\rm KL, oct} \sim 25 \ (140)$~yr for the BHNS (NSNS) system. 
A comparison with the actual evolution reveals that the analytical estimates give values within a factor of a few compared with those 
inferred by the oscillations in Figures~\ref{fig:BHNS_evolution_example} and~\ref{fig:NSNS_evolution_example}.

For both the simulated binaries, the octupole terms result to be quite relevant in the secular evolution, 
especially in the BHNS case. Indeed, a lower inner mass ratio $q$ enhances the strength of the octupole correction 
and reduce the associated oscillation time-scale, as it depends on the difference $m_1-m_2$ (see, e.g. 
equations~\ref{eq:KL octupole time scale} and~\ref{eq:C3}). Therefore, in addition to the reduced merger 
time-scale due to the higher mass with respect to the NSNS case, the lower mass ratio of the BHNS binary 
produces a much shorter octupole time-scale, which provides the possibility for the binary to reach a maximum in the eccentricity more frequently.
Finally, the case of the NSNS binary, reported in Figure~\ref{fig:NSNS_evolution_example}, also shows additional 
features during the evolution, in which after $t \sim 9.93\times 10^3$~yr, a sharp change in the oscillation 
pattern is evident. This is due to the octupole terms that cause a switch from the libration regime
(i.e. oscillation around $g_1 = \pi/2$) to the circulation regime (i.e. monotonic increase of $g_1$ in the range 
$[0,2 \pi]$) of the inner argument of pericentre \citep[see discussion in][]{Blaes_et_al_2002}, on a time-scale of a few times the octupole 
time-scale. In the latter regime, the minimum eccentricity 
is higher, which produces slightly more efficient GW emission.

Figures~\ref{fig:BHNS_evolution_example} and \ref{fig:NSNS_evolution_example} show how the features of the KL mechanism change when mass and mass ratio of the inner binary vary.  We take the converse approach in Appendix~\ref{app:parameter_exploration}, where we report a systematic exploration of the parameter space through a selected grid. We explore a few representative cases, both with NSNS and BHNS as inner binaries. We fix the masses of the inner component and vary all the other parameters that characterise the triplet. From our analysis, the most important parameters for the KL efficiency are the outer semi-major axis and the relative inclination. We address the interested reader to Appendix~\ref{app:parameter_exploration} for full details.     

%%%%%%%%%%%%%%%%%%%%%%%%%%%%%%%%%%%%%%%%%%%%%%%%%%%%%%%%%%%%%%%%%%%%%%
\section{Coalescence time-scale for stellar triplet distributions}
\label{sec:distributions}

%%%%%%%%%%%%%%%%%%%%%%%%%%%%%%%%%%%%%%%%
\begin{table*}
	\centering
	\caption{{\it Top}: Summary of the distributions applied to produce the population of triple systems discussed in Section~\ref{sec:distributions}.
		{\it Bottom}: Summary of the results obtained from the above populations. $S$, $P$, and $U$ represent the number of stable non-processing, precessing,
		and unstable triple system in each population, respectively. $X_{\rm GW,8}$ is the number of system of type $X$ whose inner binary has a GW-coalescence time-scale shorter than $10^8~{\rm yr}$ without considering the third body perturbation, whereas $S_{\rm M,8}$ is the number of triple stable, non-precessing systems whose merger time-scale is shorter than
		$10^8~{\rm yr}$. The comparison between the last two rows shows the boosting effect of triple interactions.}
	\begin{tabular}{ c | c | c | c | c }
		\hline
		%& & & & \\
		&    NSNS, case A & NSNS, case B  & BHNS, case A  &  BHNS, case B  \\
		%& & & & \\ 
		\hline \hline                 
		%& \multicolumn{4}{c}{~}      \\                                                  
		& \multicolumn{4}{c}{Distributions}    \\ \hline \hline
		%& \multicolumn{4}{c}{~}      \\   \hline                                   
		$g_1$        &  \multicolumn{4}{|c}{uniform in $[0,2 \pi]$}                \\ %\hline
		$g_2$        &  \multicolumn{4}{|c}{uniform in $[0,\pi]$}                  \\ %\hline
		$m_3~[\msun]$    &  \multicolumn{4}{|c}{Salpeter power law (slope -2.3), in $[3,15]$}           \\ %\hline
		$e_1$        &  \multicolumn{4}{|c}{uniform in $[0,1]$}                          \\ %\hline
		$e_2$        &  \multicolumn{4}{|c}{linear in $[0,1]$}                          \\ %\hline
		$a_2~[{\rm AU}]$ &  \multicolumn{4}{|c}{linear in $[0.03,10]$}                       \\ %\hline
		$\cos{i}$        &  \multicolumn{4}{|c}{uniform in $[-1,1]$}                         \\ \hline
		$m_1~[\msun]$    &  \multicolumn{2}{|c}{Gaussian in $[1.0,2.4]$}   &  \multicolumn{2}{c}{Gaussian in $[5.0,30]$}                     \\
		&  \multicolumn{2}{|c}{$\langle m_1 \rangle = 1.4$ , $\sigma= 0.13$}        &  \multicolumn{2}{c}{$\langle m_1 \rangle = 8$, $\sigma= 0.42$}   \\ %\hline
		$m_2~[\msun]$    &  \multicolumn{2}{|c}{Gaussian in $[1.0,2.4]$}   &  \multicolumn{2}{c}{Gaussian in $[1.0,2.4]$}                     \\
		&  \multicolumn{2}{|c}{$\langle m_2 \rangle = 1.4$, $\sigma= 0.13$}        &  \multicolumn{2}{c}{$\langle m_2 \rangle = 1.80$ , $\sigma=0.17$} \\ \hline
		\multirow{2}{*}{$a_1~[{\rm AU}]$} &  unif. in  & unif. in & unif. in  & unif. in  \\
		 &  $[0.003,0.3]$  & $\log_{10}{[0.003,0.3]}$ & $[0.003,0.3]$  & $\log_{10}{[0.003,0.3]}$ \\
		\hline \hline
		%& \multicolumn{4}{|c}{~} \\
		& \multicolumn{4}{|c}{Results}      \\ \hline \hline
		%& \multicolumn{4}{|c}{~}\\ \hline                
		$N = S + P + U$                  &  3346     & 3897    & 3297   & 5123 \\ \hline                  
		$S/N$                            &  0.5977   & 0.5132  & 0.6066 & 0.3904 \\ 
		$S_{\rm M,8}/N$                  &  0.0607   & 0.0426  & 0.0874 & 0.0509  \\
		$S_{\rm GW,8}/N$                 &  0.0093   & 0.0159  & 0.0173 & 0.0189   \\ \hline
		$P/N$                            &  0.0511   & 0.2969  & 0.1110 & 0.4540 \\
		$P_{\rm GW,8}/N$                 &  0.0254   & 0.1499  & 0.0658 & 0.3475 \\ \hline
		$U/N$                            &  0.3512   & 0.1899  & 0.2824 & 0.1556  \\
		$U_{\rm GW,8}/N$                 &  0.0036   & 0.0100  & 0.0103 & 0.0197  \\ \hline
		$(S_{\rm M,8}+P_{\rm GW,8})/N$   &  0.0861   & 0.1925  & 0.1532 & 0.3984  \\
		$(S+P+U)_{\rm GW,8}/N$                 &  0.0383   & 0.1758  & 0.0934 & 0.3861  \\
		\hline \hline
	\end{tabular}
	\label{tab: population distributions}
\end{table*}
%%%%%%%%%%%%%%%%%%%%%%%%%%%%%%%%%%%%%%%%

%%%%%%%%%%%%%%%%%%%%%%%%%%%%%%%%%%%%%%%%
\begin{figure*}
	\begin{minipage}[c]{0.47\textwidth}
		\centering
		\includegraphics[scale=0.41]{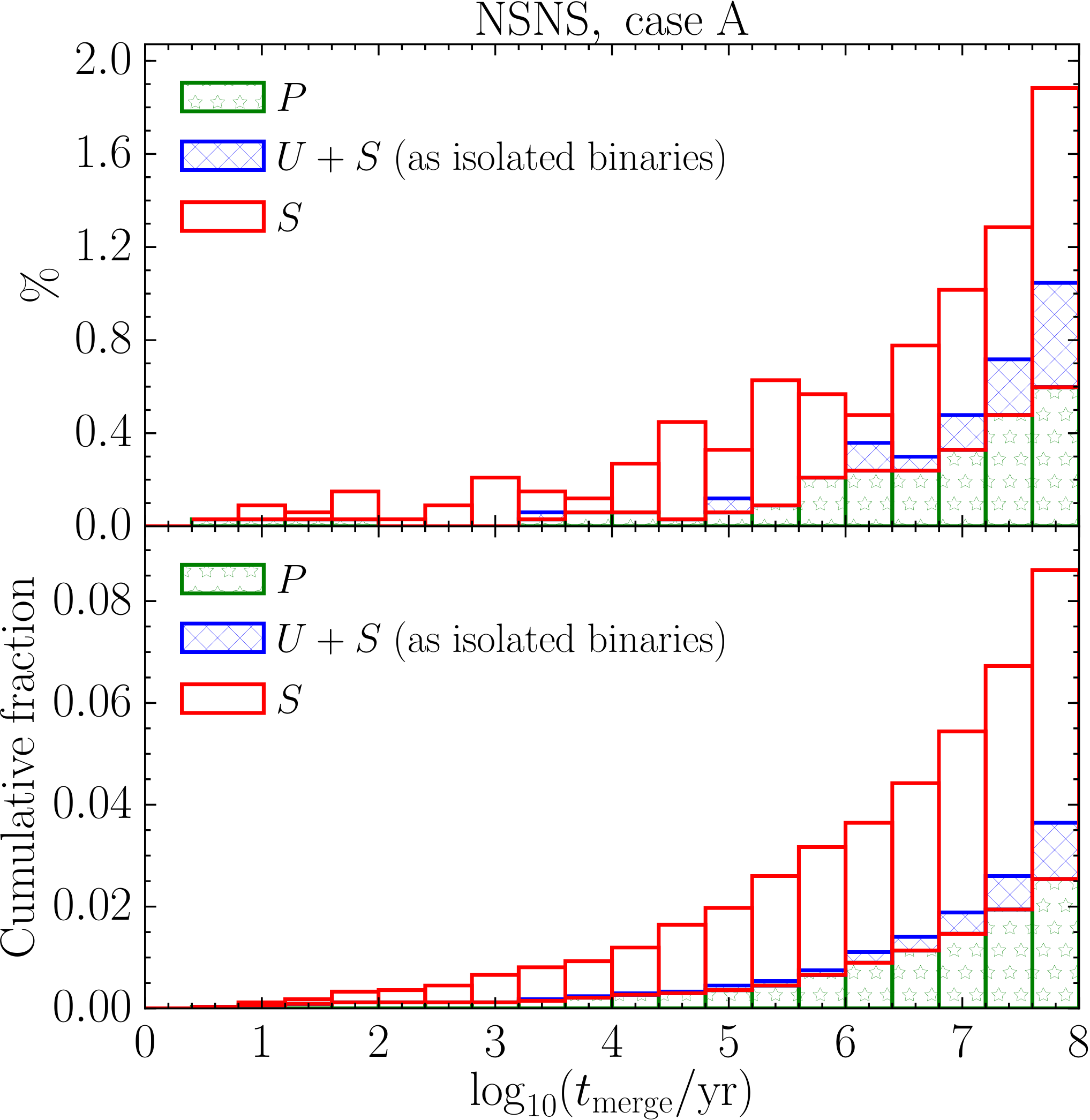}
	\end{minipage}
	\ \hspace{3mm} \
	\begin{minipage}[c]{0.47\textwidth}
		\centering
		\includegraphics[scale=0.41]{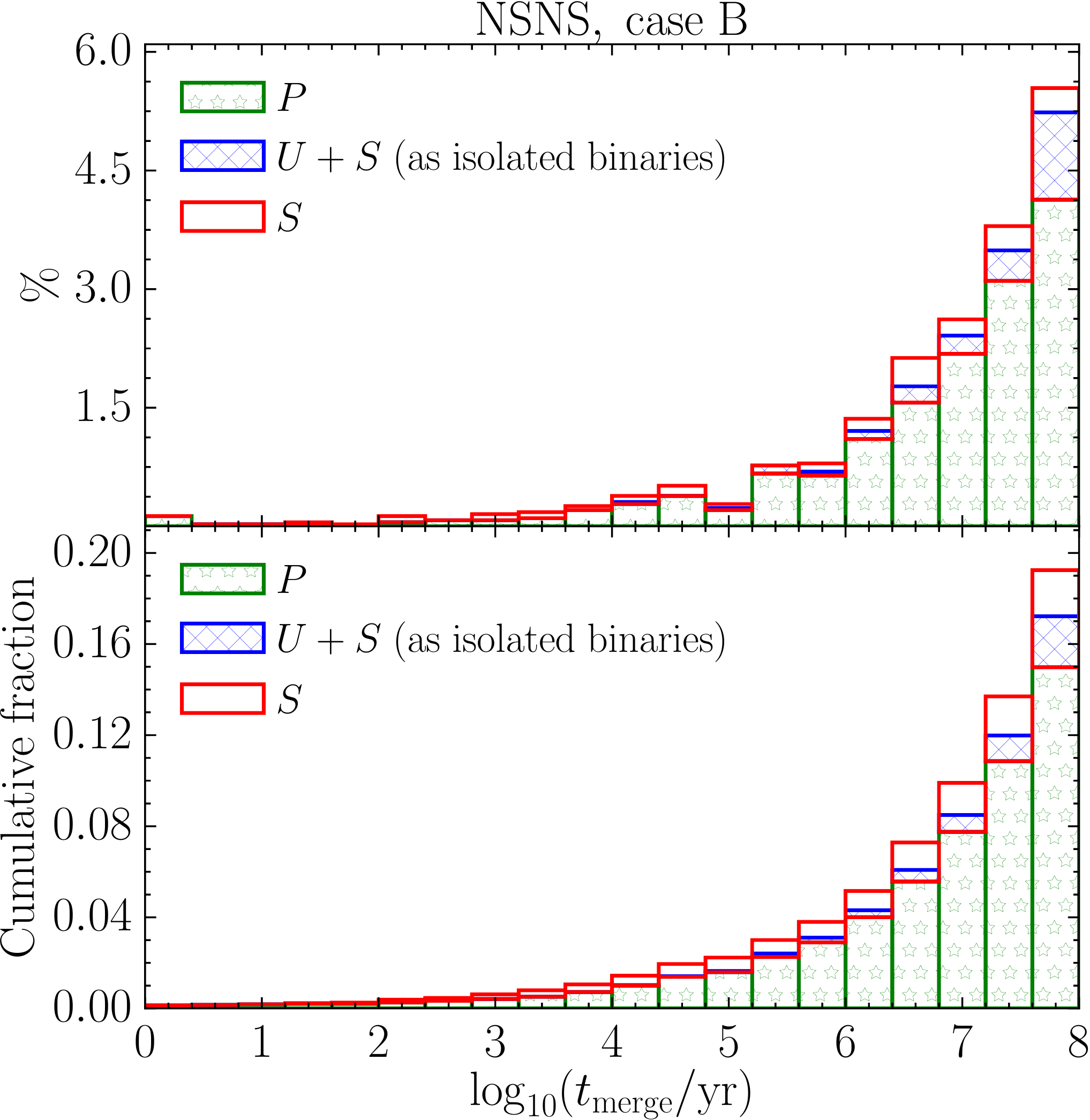}
	\end{minipage}
	\caption{Comparison of the distributions of the merger time-scale below $10^8 \ {\rm yr}$ 
		for NSNS binaries in triplets ($t_{\rm merge}$) and for the same binaries assumed as isolated
		(i.e. $t_{\rm GW}$). Details of the distributions are specified in Table~\ref{tab: population distributions}. 
		Green bars (filled with stars) include triplets for which the relativistic precession of the inner binary strongly 
		inhibits the effect of secular effects. For these systems, we assume $t_{\rm merge} \approx t_{\rm GW}$. 
		Blue bars (filled with lines) include $t_{\rm GW}$ of the inner binary both for hierarchical, non precessing triplets 
		and unstable triplets. Red bars (unfilled) contain hierarchical, non precessing systems considered as triplets. 
		{\it Left panels}: initial inner binary distribution uniform in $a_1$. 
		{\it Right panels}: initial inner binary distribution uniform in $\log_{10}{(a_1)}$.
		{\it Upper panels}: percentage of runs.
		{\it Lower panels}: cumulative fraction of runs.}
	\label{fig:NSNS_histo_a1_log_lin}
\end{figure*}
%%%%%%%%%%%%%%%%%%%%%%%%%%%%%%%%%%%%%%%%
%%%%%%%%%%%%%%%%%%%%%%%%%%%%%%%%%%%%%%%%
\begin{figure*}
	\begin{minipage}[c]{0.47\textwidth}
		\centering
		\includegraphics[scale=0.41]{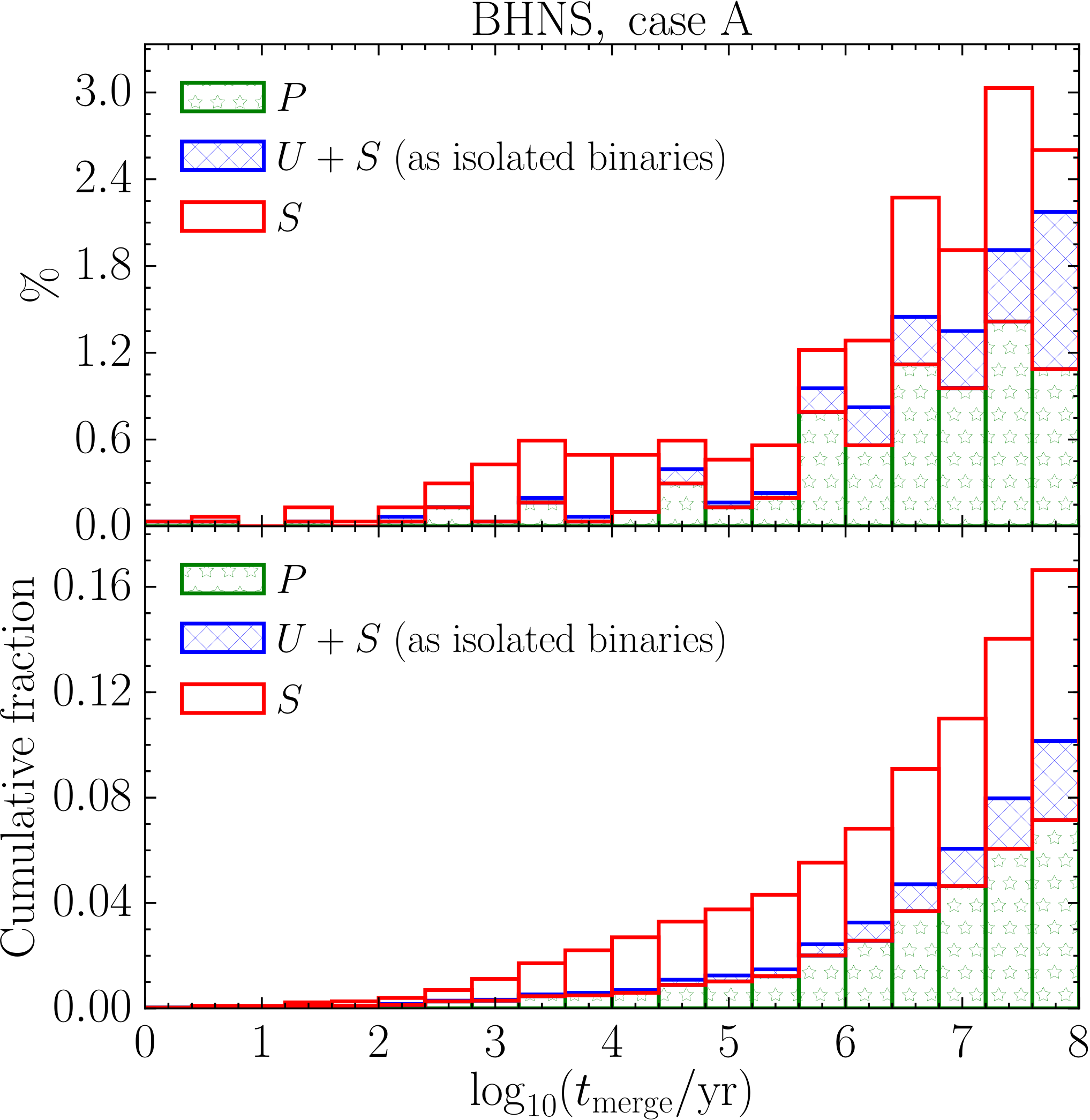}
	\end{minipage}
	\ \hspace{3mm} \
	\begin{minipage}[c]{0.47\textwidth}
		\centering
		\includegraphics[scale=0.41]{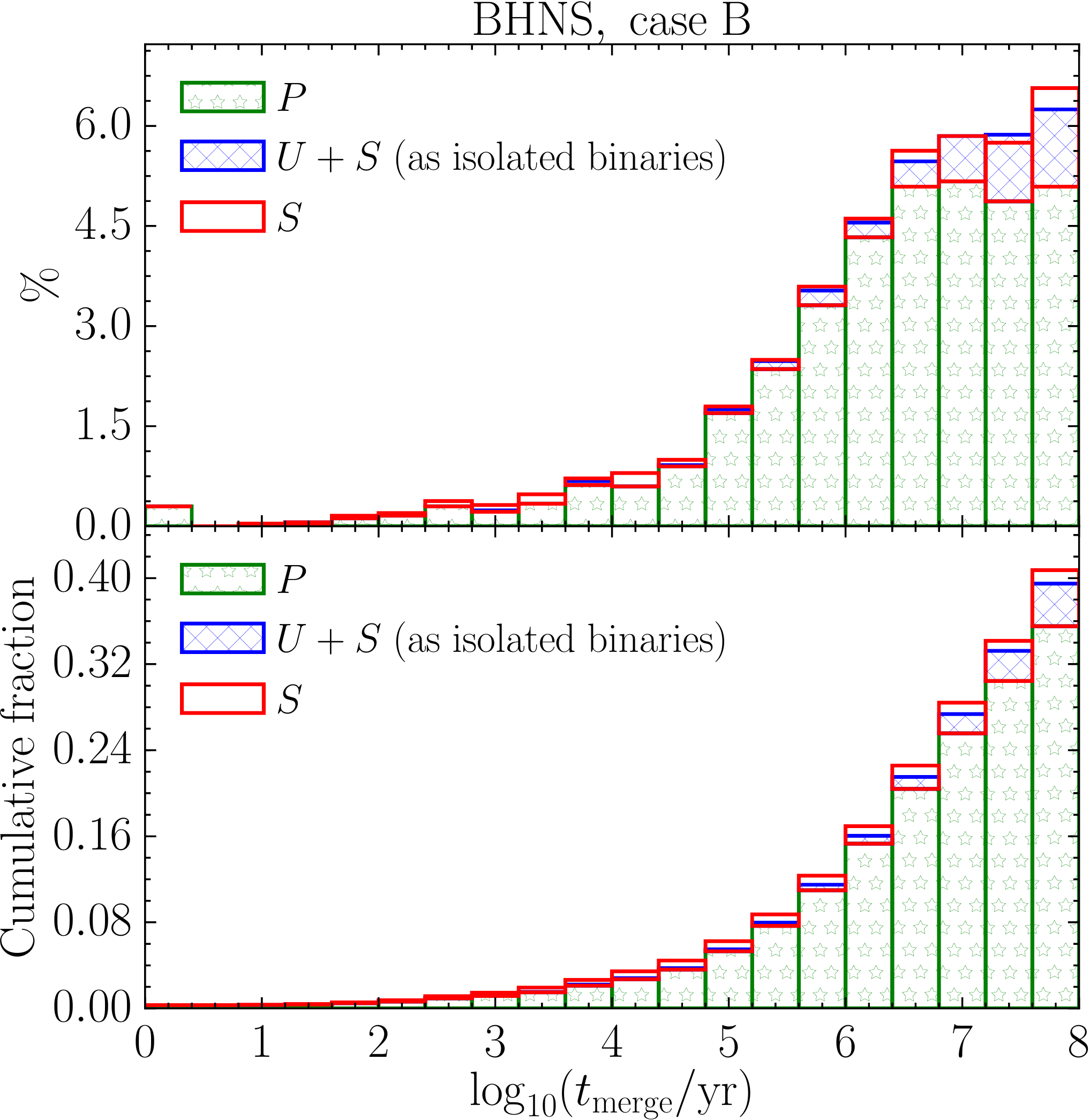} 
	\end{minipage}
	\caption{Same as Figure~\ref{fig:NSNS_histo_a1_log_lin}, but for BHNS inner binaries.}
	\label{fig:BHNS_histo_a1_log_lin} 
\end{figure*}
%%%%%%%%%%%%%%%%%%%%%%%%%%%%%%%%%%%%%%%%

To test the impact of triple system dynamics on the merger time-scale of a population of compact binaries,
we generate different populations of triplets, all characterized by an inner compact binary and an orbiting outer star.
We consider separately NSNS and BHNS inner binaries, and we vary the distribution of the inner semi-major axis between
two cases, for a total of four different populations.
The initial conditions characterizing each triplet are generated through Monte Carlo sampling.
A set of distributions is common to all populations and it includes:

%%%%%%%%%%%%%%%%%%%%%%%%%%%%%%%%%%%%%%%%
\begin{itemize}
	\item for $g_1$ and $g_2$, uniform distributions between 0 and $2\pi$, and between 0 and $\pi$, respectively.
	The precise value of the two arguments of periastron depends on the details of the triplet formation. 
	We assume isotropy and no correlation between the formation of the inner and outer binary. Moreover,
	we employ the symmetry presented at the end of Section~\ref{sec:secular evolution} to halve the range of $g_2$;
	\item for $i$, a uniform distribution in $\cos{i}$ between $-1$ and $1$, which is equivalent to an isotropic probability for the direction
	of $\mathbf{G}_2$ with respect to $\mathbf{G}_1$;
	\item for $m_3$, a Salpeter \citep{Salpeter1955} distribution with slope -2.3 between $3$ and $15~\msun$ (see the discussion of $m_3$ in Section~\ref{sec:estimates});
	\item for $e_1$, a uniform distribution between 0 and 1, because the observed NSNS binaries have a broad distribution and the actual
	value of $e_1$ does not have a strong impact on the evolution of the triplet; 
	\item for $a_2$ and $e_2$, a linear distribution, i.e. $f(x) \propto x$, between $3 \times 10^{-2}$ and $ 10 \ {\rm AU}$, and
	between 0 and 1, respectively. This kind of distribution is expected to be appropriate when triplets form dynamically \citep{Heggie:1975}.
\end{itemize}
%%%%%%%%%%%%%%%%%%%%%%%%%%%%%%%%%%%%%%%%
For the NS masses in NSNS (BHNS) inner binaries, we consider $1.0 \leq m_{\rm NS} \leq 2.4~\msun$ 
and we assume a Gaussian distribution centred around $1.4~\msun$ ($1.8~\msun$), 
with standard deviation $0.13~\msun \ (0.18~\msun)$ \citep{Dominik2012}.
For the BH masses in BHNS inner binaries, we take $ 5 \leq m_{\rm BH} \leq 30~\msun $,
and we also assume a Gaussian distribution centred around $ 8~\msun$,
with standard deviation $0.42~\msun$ \citep{Dominik2012}.
Finally, for the inner binary separation, we consider two possibilities: case A, a distribution uniform in log$_{10}(a_1)$; 
and case B, a distribution uniform in $a_1$.
The orbital parameter distributions used to generate the triplets are summarized in the upper part of Table~\ref{tab: population distributions}.

For each population, we randomly generate $N$ triple systems and we distinguish among precessing ($P$), unstable ($U$), and stable, non-precessing ($S$) 
systems according to equations~(\ref{eq:stability criterium}) and~(\ref{eq:precession criterium}). Clearly, $N = P+U+S $.
We produce $N$ triple systems such that $ S =  2000$. For the precessing systems, the coalescence time is assumed to be $t_{\rm GW}$, independent of the 
presence of the third external body. For unstable systems, we assume that the inner binary is always disrupted by the presence of the third body,
which probably ejects the lighter compact object (i.e. the NS) from the innermost binary\footnote{We verified our assumption by simulating the 
	triplet evolution of a large sub-sample of the unstable systems
	using the code developed in \citet{Bonetti2016}.}. 
Thus, these systems will never lead to a compact binary coalescence 
when considered as part of a triple system.
Finally, for the stable, non-precessing triples, we compute the merger time by integrating the equations of motion (cf. equations~\ref{eq:da1dt}--\ref{eq:dhdt}).
We compare the distribution of the merger times for the triple systems with the distribution of 
$t_{\rm GW}$ for the $N$ inner binaries (i.e. always neglecting the effect of the third body). We normalize both distributions to $N$ to find
the fraction of inner binaries that coalesce within $10^8~{\rm yr}$, with and without the presence of the third body.
In the lower part of Table~\ref{tab: population distributions}, we summarize the results obtained for our four populations.

In Figure~\ref{fig:NSNS_histo_a1_log_lin}, we show our results for the NSNS distributions, both in the case of a uniform distribution in 
$a_1$ (left panel, case A) and in $\log_{10}(a_1)$ (right panel, case B).
The precessing triplets merging within $t_{\rm merge}$ are common both to the triple and binary distributions
(green star bars). The KL mechanism leads to an increase of the merger rate (red empty bars), even when considering the systematic disruption of the inner binary when part of unstable triple systems.
In case A, the uniform distribution of the inner semi-major axis, combined with the linear distribution of the outer semi-major axis, favours the presence of
stable, non-precessing triplets ($\sim 60\%$ of the cases). The few precessing systems are characterized by tight inner binaries, which coalesce within $10^8$~yr
in $\sim 50\%$ of the cases. The remaining unstable systems have rather large initial $a_1$ and only a very small fraction of their inner compact binaries 
($\sim$ 1\%) would merge as isolated binaries.
Overall, only $3.8 \%$ of the inner systems of this population would coalesce within $10^8$~yr as isolated binaries.
For stable, non-precessing systems, the KL mechanism causes a fast merger of the inner binary in one case out of ten, which is increased by a factor of 6.5 compared
with the fraction of merging isolated binaries. Considering the whole population,
the number of systems coalescing within $10^8$~yr as triplets has increased by a factor 2.25, to $8.6\%$ of the population.

The $\log_{10}$-uniform distribution of inner semi-major axis used in case B produces qualitatively different results. The presence of a much larger number of
tight inner binaries increases the number of precessing systems at the expense of the unstable and, less severely, of the
stable, non-precessing systems. Also in this case, more than 50\% of the inner binaries contained inside the precessing triplets will coalesce anyway within $10^8$~yr.
The KL mechanism increases the number of fast coalescences in stable, non-precessing systems by a factor of 2.7. However, due to the dominant
presence of tight, precessing systems, the total fraction of fast coalescing systems increases only from 17.6\% to 19.25\%, when passing from isolated binaries
to triplets.
The temporal distributions reported in Figure~\ref{fig:NSNS_histo_a1_log_lin} suggest also that the number of coalescing systems increases with $t_{\rm merge}$
for all system types. However, the increase is more pronounced for precessing and unstable systems. Thus, the KL mechanism is very efficient in increasing
the number of mergers on extremely short time-scales ($t_{\rm merge} < 10^5$~yr).

The results obtained for the BHNS inner binary cases are reported in Figure~\ref{fig:BHNS_histo_a1_log_lin}, both for a uniform distribution in 
$a_1$ (left panel, case A) and in $\log_{10}(a_1)$ (right panel, case B). The qualitative behaviour of the NSNS populations described above is also 
valid in the case of BHNS populations.
The presence of a stellar-mass BH in the inner binary increases $m_1 + m_2$, leading
to a more efficient GW emission and a significantly shorter $t_{\rm GW}$, since $t_{\rm GW} \propto [(m_1 + m_2)m_1 m_2]^{-1}$ (see equation~\ref{eq:t_gw}). 
It also increases the stability of triple systems (see equation~\ref{eq:stability criterium}), but favours the relativistic precession 
of the inner binary (see equation~\ref{eq:precession criterium}). Moreover, the combination with the $a_1^{4/3}$ dependence in equation~(\ref{eq:precession criterium}) makes
the occurrence of precession even more pronounced, moving from case A to case B.
The more massive inner binary makes the KL resonance induced by the third body less efficient (this is visible, for example, 
on the longer time-scale for the dominant quadrupole oscillations; see equation~\ref{eq:KL quadrupole time scale}).
On the other hand, the larger mass difference potentially increases the importance of octupole modulation
(see Section~\ref{sec:inner binary evolution}).
For a uniform distribution in $a_1$ (case A), the largest contribution to the number of inner binaries that would coalesce as isolated binaries is
provided by tight precessing systems (6.58\% of the whole population). The KL mechanism increases the number of compact binaries that have a fast coalescence 
in stable, non-precessing systems by a factor of 5, and up to 8.74\% of the population, i.e.
in a way similar to what reported for the NSNS population of case A. In total, the fraction of BHNS binaries that coalesce within $10^8$~yr
has increased from 9.34\% as isolated binaries to 15.3\% as inner binaries of a population of triplets. The larger absolute values, compared with the NSNS
population, are simply due to the more efficient GW emission, while the impact of the KL mechanism has slightly decreased, due to the more massive inner binary.
The even more reduced impact of the KL mechanism on the fraction of the fast coalescing, stable, non-precessing systems becomes marginal 
in case B of the BHNS population.
For the latter, the largest fraction ($\gtrsim 38 \%$) of fast coalescing system is represented by precessing systems, which merge within $10^8$~yr in 
$\sim 75\%$ of the cases. 

%%%%%%%%%%%%%%%%%%%%%%%%%%%%%%%%%%%%%%%%%%%%%%%%%%%%%%%%%%%%%%%%%%%%%%
\section{Discussion and conclusions}
\label{sec:conclusions}

In this work, we have analysed the impact of the KL mechanism on the merger rate of compact binaries (both BHNS and NSNS) in the early stage of the cosmological evolution. 
Our investigations are motivated by the observation of $r$-process elements in old, metal poor stars, which demands the occurrence of $r$-process nucleosynthesis
for ${\rm [Fe/H]} < -3$ (corresponding to a delay of $\sim 10^8$~yr after the birth of the first stars in the case of efficient elemental mixing in the galactic interstellar medium).
We have verified that the KL mechanism can, under certain conditions, be important in shaping the merger rate of compact binaries.
Our results confirm previous findings of \cite{Thompson2011}, who showed that the KL mechanism can be relevant in increasing the merger
rate of compact binaries on time-scales comparable to the Hubble time. However, we have specialised to the case of fast ($\sim 10^8$~yr) mergers, 
for which we have found the following.

On the one hand, if the main compact binary formation channel favours the occurrence of tight compact systems (for instance with $a_1$ distributed uniformly in logarithm), then the influence of the KL mechanism is negligible because the merger fraction increases by only a few percent.
This is due to the stronger relativistic precession that characterises tighter binaries and destroys the KL resonance. 
However, in this scenario, given the smaller average inner separations, a significant fraction of binaries efficiently merges in short time-scales 
without any external influence \citep[see, e.g.][]{Beniamini.etal:2016}.

On the other hand, if the distribution of the semi-major axes favour the formation of wider inner compact binaries, then the merger rate of NSNS and BHNS binaries can be increased 
up to a factor of two because of secular triple interactions. 
Since in this situation the fraction of tight binaries that efficiently merge in less than 100 Myr is low (only a few percent), triple interactions should not be neglected and
the KL mechanism can be crucial, if compact binary mergers are the main site for the production of $r$-process elements in the early Universe. 

A remarkable feature of the enhanced CBM rate due to the KL mechanism is the occurrence of
ultra-fast merger events ($\lesssim 10~{\rm Myr}$). Such a reduced merger time-scale could be crucial to
explain the observed abundances in $r$-process enriched ultra-faint dwarf galaxies (e.g. Reticulum II)
with a single CBM event \citep{Safarzadeh.Scannapieco:2017}. Indeed, the shallow potential well 
of the ultra-faint dwarf halos, combined with the potentially large natal kick of compact binaries, 
requires ultra-fast mergers so that the merger does not happen outside the galaxy and 
to prevent interstellar medium enrichment (see, e.g. \citealp{Safarzadeh.Cote:2017}, but see also \citealp{Beniamini.etal:2016},
for the possible impact of low natal-kick, tight binaries).

We have performed our study under the assumption of secular evolution, up to octupole-order KL equations. However, we cannot exclude that 
the inclusion of higher-order effects or the study of non-hierarchical situations could be relevant, at least for a part of the wide parameter
space. A more detailed study, employing direct integration schemes, will be the subject of forthcoming investigations.

Despite the potential relevance of the KL mechanism for the merger rate of compact binaries, several questions concerning the 
formation rate and properties of triple systems remain unanswered. A first question is whether hierarchical triple systems
can easily form and if they are frequent enough. The total fraction of massive stars that are located in multiple systems 
is $\gtrsim 80\% $ \citep{Duchne2013}, with a significant portion ($\sim 10\%$) in triple or even quadruple systems \citep[see][and references therein]{Belczynski2014}.
Recent hydrodynamical simulations of primordial star formation predict that the collapse of metal-free
clouds of H and He likely forms multiple systems \citep{Stacy2010,Clark2011,Girichidis2012}.
Moreover, the initial mass function for metal-free stars can differ significantly from what we observe at later epochs 
\citep[e.g.][and references therein]{Hartwig.etal:2015} and increase
the presence of more stable high-mass tertiary components, for which we expect the KL mechanism to be more efficient.
A second question concerns the places and the channels through which these systems can be born.
Triple systems can form either in GCs or in the GF.
The formation probability is larger in GCs, because they are denser stellar environments.
Indeed, the formation of compact binaries in high-redshift GCs can already enhance the merger rate in the 
early Universe \citep{Ramirez-Ruiz.etal:2015}. 
However, in a Milky Way-like galaxy, only $\sim10^7$ out of $\sim10^{11}$ stars are located in GCs. Thus, triple systems in the GF are also relevant.
A first channel to produce hierarchical triple systems is in-situ formation. This can happen both in GCs and in the GF.
For fixed energy and angular momentum, there is more phase-space in which the lighter object is outside. In this case,
the inner system can evolve in a compact binary, while the outer body stays an ordinary star.
Although the inner and outer angular momenta are initially aligned, asphericity in the SN explosions of the inner binary 
can lead more easily to misaligned configurations.
Another channel is the dynamical formation of a triple system from the
capture of a third body by a compact binary.  However, because in the
Newtonian point-mass approximation the orbits are time-reversible, the
formation of a stable hierarchical triple is only possible if energy
can be dissipated, e.g. via tidal effects or the emission of
GWs \citep[see][]{Bailyn1989}. Finally, an other feasible channel is the interaction between a compact binary and another wider binary, which can trigger the
ejection of the lighter component of the latter and the formation of a stable triplet.
Dynamical channels are expected to be more likely in GCs where perturbations
due to the global distribution of stars are expected to be more relevant for wider, triple systems than for binaries.
If these perturbations induce changes in the relative inclination, the probability to access the KL-favourable range
could be increased \citep[see, e.g.][]{VanLandingham2016}. If they trigger instabilities or exchanges, this could lead to
a shrinking of the semi-major axis or to an increase of the eccentricity of the semi-major axis.

%%%%%%%%%%%%%%%%%%%%%%%%%%%%%%%%%%%%%%%%%%%%%%%%%%%%%%%%%%%%%%%%%%%%%%
\begin{acknowledgements}
The authors thank A. Arcones, M. Pignatari, M. Safarzadeh, F.-K. Thielemann, and B. Wehmeyer for useful discussions.
MB and MD acknowledge the CINECA award under the ISCRA initiative,
for the availability of high performance computing resources and support.
AP acknowledges support from the Helmholtz-University Investigator grant No. VH-NG-825, and
from the INFN project ``High Performance data Network'' funded by CIPE.
This work was supported by a grant from the Swiss National Supercomputing Centre (CSCS) under 
project ID 667. AP thanks also the GSI Helmholtzzentrum f\"ur Schwerionenforschung GmbH 
for the usage of computational resources. 
PRC acknowledges support by the Tomalla foundation.
\end{acknowledgements}

%%%%%%%%%%%%%%%%%%%%%%%%%%%%%%%%%%%%%%%%%%%%%%%%%%%%%%%%%%%%%%%%%%%%%%
%%%%%%%%%%%%%%%%%%%%%%%%%%%%%%%%%%%%%%%%%%%%%%%%%%%%%%%%%%%%%%%%%%%%%%

\bibliographystyle{pasa-mnras}
\bibliography{biblio}

%%%%%%%%%%%%%%%%%%%%%%%%%%%%%%%%%%%%%%%%%%%%%%%%%%%%%%%%%%%%%%%%%%%%%%
%%%%%%%%%%%%%%%%%%%%%%%%%%%%%%%%%%%%%%%%%%%%%%%%%%%%%%%%%%%%%%%%%%%%%%

\appendix
\section{Extensive parameter exploration}
\label{app:parameter_exploration}

%%%%%%%%%%%%%%%%%%%%%%%%%%%%%%%%%%%%%%%%
\begin{table*}
	\centering
	\caption{Parameter space sampling.}
	\begin{tabular}{ c | c | c | c | c | c }
		\hline\hline
		\multicolumn{6}{c}{Parameter space} \\
		\hline
		& NSNS, I & NSNS, II & BHNS, I &  BHNS, II & BHNS, III\\
		\hline
		$m_1~[\msun]$    &  1.3 & 1.6   &  7.5 & 9.0 & 15 \\
		$m_2~[\msun]$    &  1.1 & 1.2   &  1.2 & 1.8 & 1.8 \\
		\hline    
		& \multicolumn{5}{c}{~} \\          
		$(g_1,g_2)~[\rm deg]$ & \multicolumn{5}{c}{$(90^\circ,270^\circ), (180^\circ,0^\circ)$} \\
		%$g_2~[-]$        &  \multicolumn{6}{c}{$0^\circ, 270^\circ$} \\ 
		$m_3~[\msun]$    &  \multicolumn{5}{c}{$\{1,4,7,10,13,16\}$} \\ 
		$e_1$        &  \multicolumn{5}{c}{$\{0.15,0.3,0.45,0.6,0.75,0.9\}$} \\ 
		$a_1~[{\rm AU}]$ &  \multicolumn{5}{c}{$\{0.005,0.014,0.039,0.108,0.3\}$} \\ 
		$e_2$        &  \multicolumn{5}{c}{$\{0.2,0.4,0.6,0.8\}$} \\ 
		$a_2~[{\rm AU}]$ &  \multicolumn{5}{c}{$\{0.03,0.096,0.306,0.979,3.129,10\}$} \\ 
		$\cos{i}$        &  \multicolumn{5}{c}{$\{0.866,0.779,0.693,0.606,0.52,0.433,0.347,0.26,0.174,0.087,-0.087\}$} \\ 
		\hline\hline
	\end{tabular}
	\label{tab:grid_distributions}
\end{table*}
%%%%%%%%%%%%%%%%%%%%%%%%%%%%%%%%%%%%%%%%

In this appendix, we report on a broader parameter space exploration of hierarchical, non-precessing triple systems with few selected masses for the inner compact binary. The main goal of this study is to highlight which parameters are most relevant in shaping the KL efficiency, eventually causing binary coalescence. In Table~\ref{tab:grid_distributions}, we summarize the surveyed parameter space and its sampling. For the NSNS (BHNS) case, we choose two (three) different mass combinations, and for each of them two further choices of the initial inner and outer arguments of pericentre (i.e. $g_1,g_2$). For $m_3$ we choose six values in the range $[1,16] \ \rm M_{\odot}$, whereas for the inner ($e_1$) and outer ($e_2$) eccentricities we select six and four values uniformly spaced in the range $[0,1]$, respectively. The inner ($a_1$) and outer ($a_2$) semi-major axes take instead five and six logarithmically spaced values from 0.005 to 0.3 AU and from 0.03 to 10 AU, respectively. Finally, we choose the relative inclination uniformly spaced in the cosine from $30^\circ$ to $85^\circ$. In addition, according to the findings of \citet{Miller2002}, we also choose to explore a single retrograde case with relative inclination of $95^\circ$. 

\begin{figure*}
	\centering
	\includegraphics[scale=0.35]{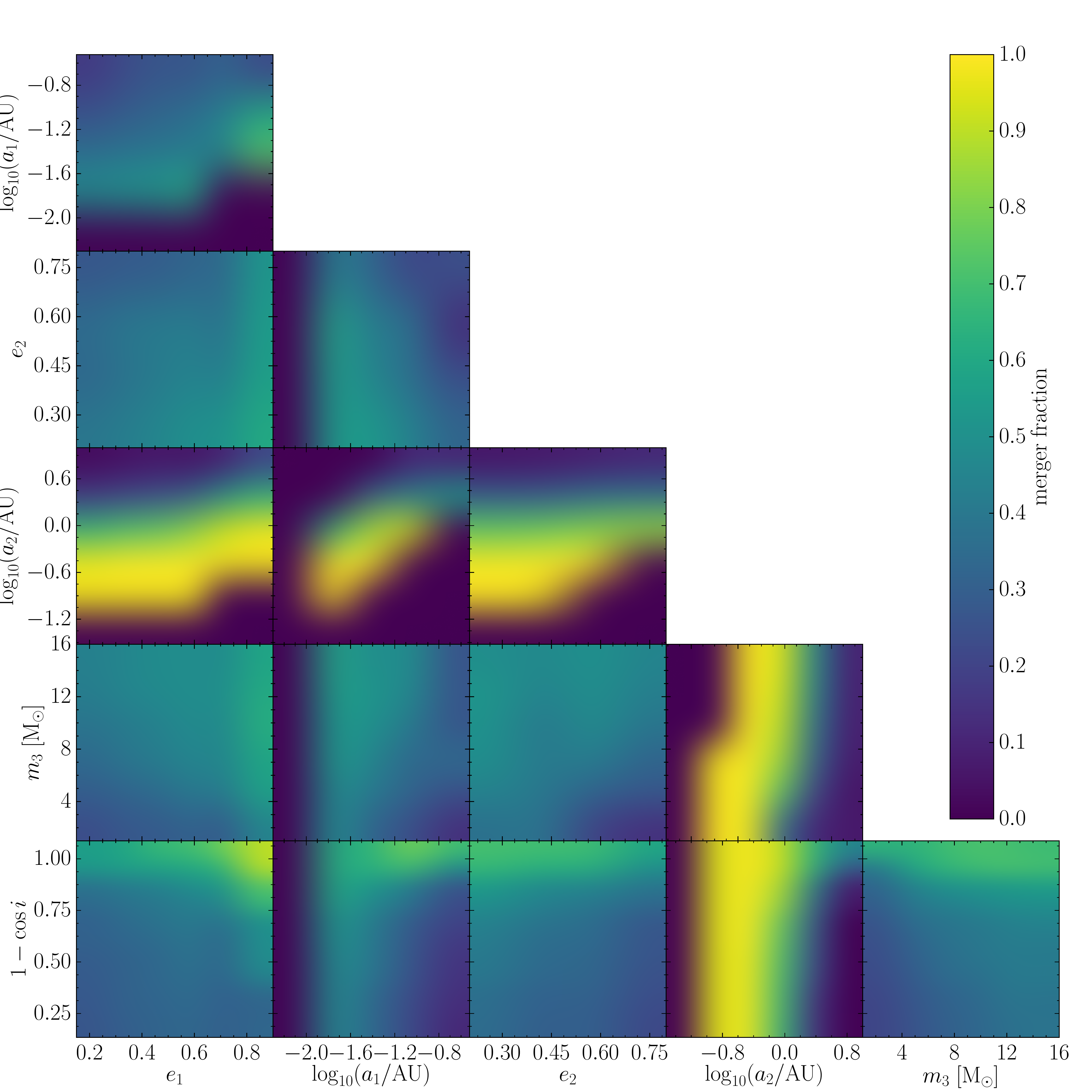}
	\caption{Merger fraction (colour-coded) as a function of various parameter pairs for the NSNS case with $m_1 = 1.6~\msun, m_2 = 1.2~\msun$, and $(g_1,g_2) = (180^\circ,0^\circ)$. Panels represent 2D slices of the merger fraction of stable non-precessing triplets that would not merge within $10^8$~yr as isolated binaries, but that do so as inner binaries of triplets because of the KL mechanism.
	We span the full range of possible combinations (see Table~\ref{tab:grid_distributions}). From the plot, the parameter $a_2$ is the most important in shaping the value of the merger fraction (cf. green/yellow areas in the plots). A relevant role is also played by the relative inclination $i$, which at values close to $90^\circ$ triggers substantial KL oscillations. 
	The sharp decreases (dark blue areas) occur instead because such points in the grid yield unstable or rapidly precessing systems, preventing or making pointless the corresponding simulations within our framework (see Section~\ref{sec:secular evolution}).}
	\label{fig:grid1}
\end{figure*}

\begin{figure*}
	\centering
	\includegraphics[scale=0.35]{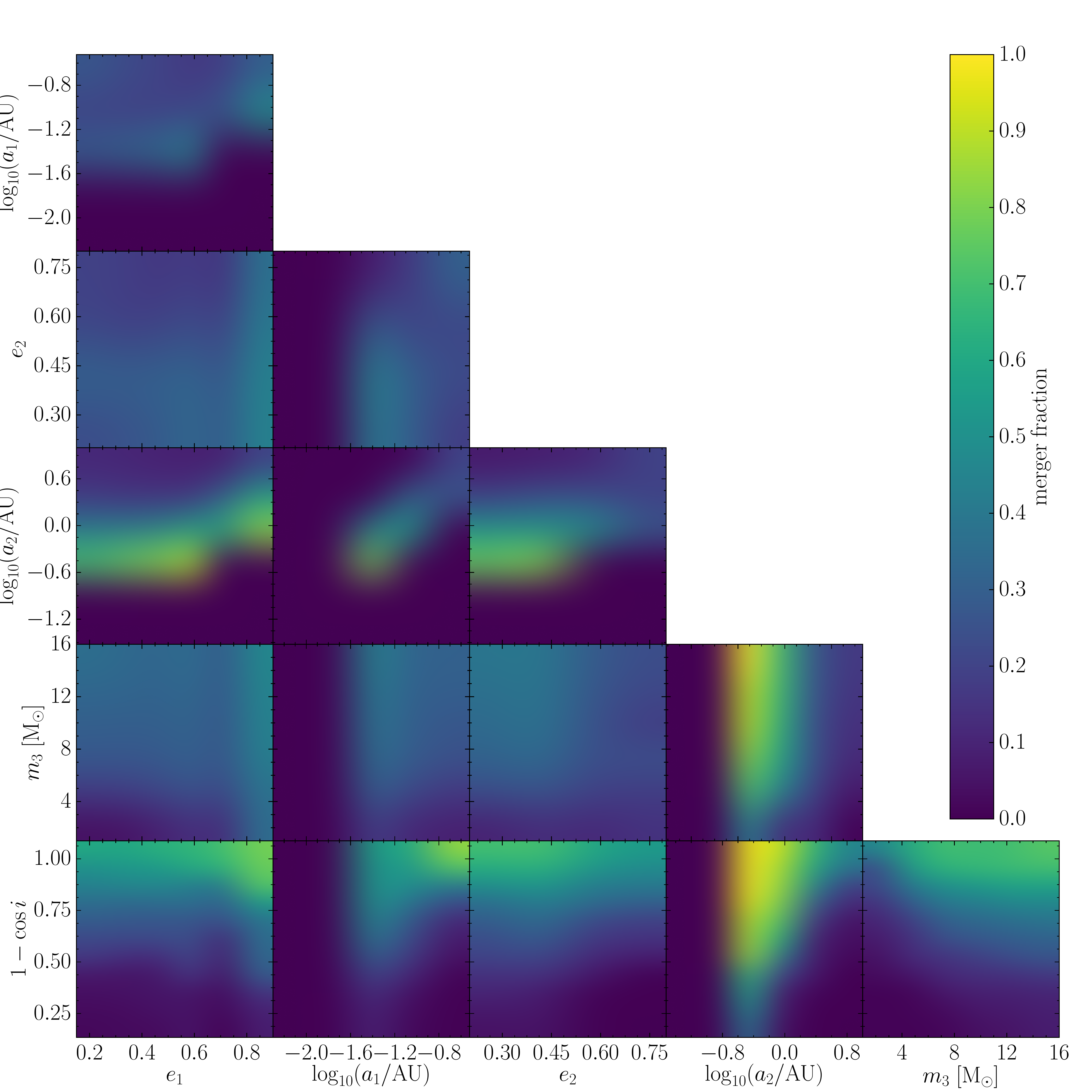}
	\caption{Same as Figure~\ref{fig:grid1}, but for the BHNS case with $m_1 = 15~\msun$ and $m_2 = 1.8~\msun$.}
	\label{fig:grid2}
\end{figure*}

\begin{figure*}
	\centering
	\includegraphics[scale=0.35]{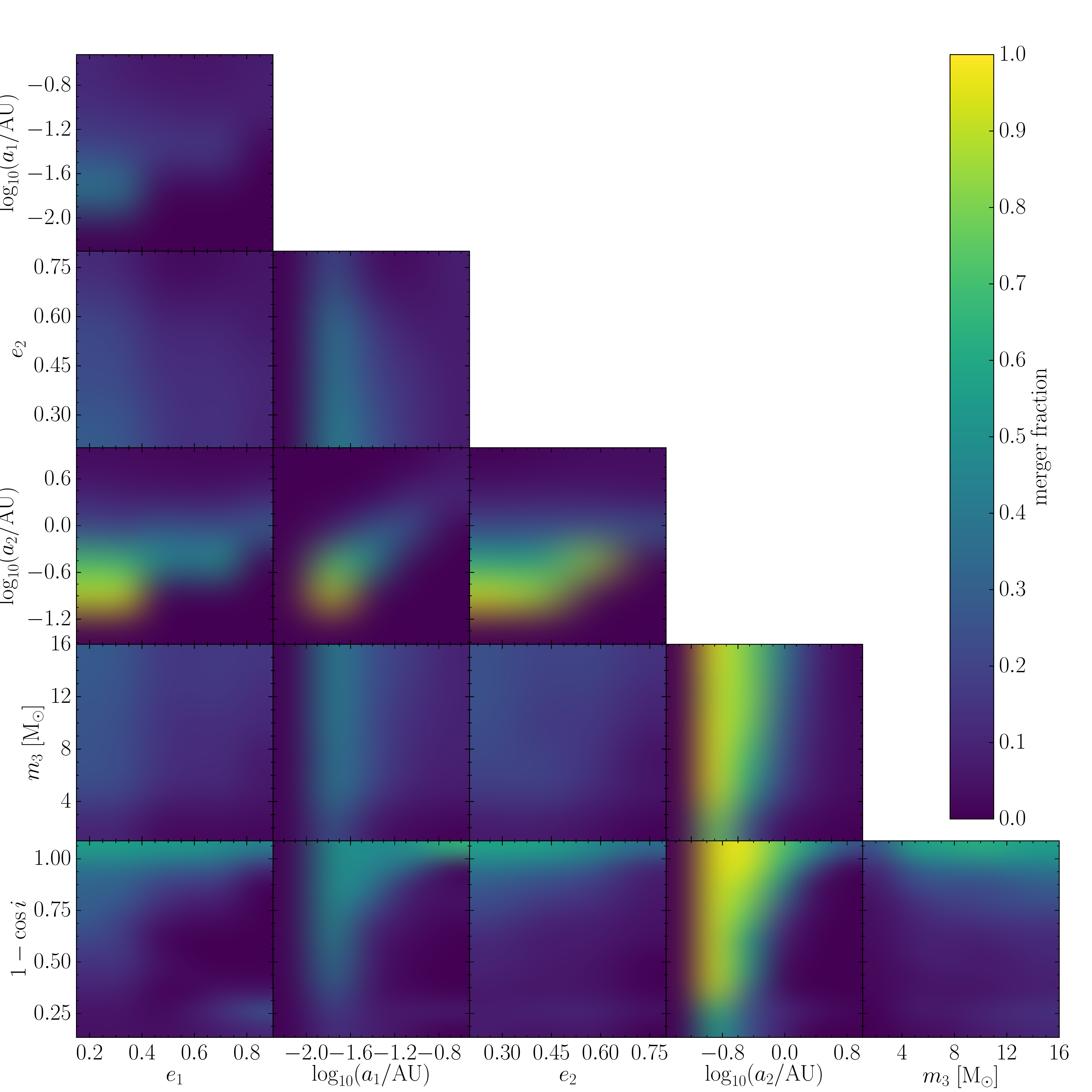}
	\caption{Same as Figure~\ref{fig:grid1}, but for the BHNS case with $m_1 = 7.5~\msun, m_2 = 1.2~\msun$, and $(g_1,g_2) = (90^\circ,270^\circ)$.}
	\label{fig:grid3}
\end{figure*}

In Figures~\ref{fig:grid1}--\ref{fig:grid3}, we report the merger fraction (colour-coded) of three representative cases (i.e. NSNS II and BHNS III with $(g_1,g_2) = (180^\circ,0^\circ)$, and BHNS II with $(g_1,g_2) = (90^\circ,270^\circ)$; see Table~\ref{tab:grid_distributions}) as a function of any possible combination ($p_1$, $p_2$) of two different grid parameters.
For every possible pair of values of $p_1$ and $p_2$, we consider the sample represented by stable and non-precessing triplets for which $t_{\rm GW} > 10^{8}~{\rm yr}$.
The merger fraction is computed as the number of grid points for which $t_{\rm merge} < 10^{8}~{\rm yr}$, normalized to the total number of points
in the sample.\footnote{We assign a merger fraction of zero also in the case when there are no stable and non-precessing triplets for a specific combination of values of $p_1$ and $p_2$.}
A merger fraction close to one implies that the KL mechanism makes the (otherwise, slowly merging) inner binary always coalesce within $10^{8}~{\rm yr}$, irrespective of all the other parameters. A merger fraction close to zero could correspond to a configuration of $p_1$ and $p_2$ for which the KL mechanism is not efficient enough, or for which stable, non-precessing systems are absent, or for which the inner binary coalesces within $10^8~{\rm yr}$ even in the absence of triple interactions.
As can be inferred from the plots, the parameter $a_2$ is the most relevant in shaping the merger fraction. Indeed, all combinations including $a_2$ show a strongly clustered pattern. The strong dependence on $a_2$ arises because the KL time-scales themselves depend on a high power of the outer semi-major axis (see equations~\ref{eq:KL quadrupole time scale} and~\ref{eq:KL octupole time scale}). Therefore, mild variations in $a_2$ lead to large changes in the KL oscillation time-scale, which in turn control how frequently the maximum inner eccentricity is reached, with its resulting copious emission of GWs. A further important role is played by the relative inclination, which leads to a high merger fraction when its value is close to $90^\circ$. In contrast, although the tertiary mass, $m_3$, can affect the oscillation time-scale, it does not seem to have a critical impact in the explored mass range. These features are common both to NSNS and BHNS systems.

A further parameter which one might expect to be important is the inner semi-major axis, $a_1$, which strongly characterises the merger time-scale of compact binaries. However, it affects the merger fraction of binaries in triple systems only marginally. The reason has to be ascribed to our exploration strategy, which here is solely directed to the assessment of the KL efficiency and not to the overall merger fraction. Indeed, a large fraction of tight inner binaries precess (see equation~\ref{eq:precession criterium}), or merge rapidly (see equation~\ref{eq:t_gw}), whereas wide inner binaries are more unstable (see equation~\ref{eq:stability criterium}). This explains the mild dependence on $a_1$ and also the sharp decreases (dark blue areas) that affect the merger fraction. The lower merger fractions visible for the BHNS cases are due to the more efficient GW emission, which increases significantly
the number of binaries that would fast coalesce also as isolated binary.

\clearpage
%%%%%%%%%%%%%%%%%%%%%%%%%%%%%%%%%%%%%%%%
%%%%%%%%%%%%%%%%%%%%%%%%%%%%%%%%%%%%%%%%
%%%%%%%%%%%%%%%%%%%%%%%%%%%%%%%%%%%%%%%%
%%%%%%%%%%%%%%%%%%%%%%%%%%%%%%%%%%%%%%%%
%%%%%%%%%%%%%%%%%%%%%%%%%%%%%%%%%%%%%%%%
%%%%%%%%%%%%%%%%%%%%%%%%%%%%%%%%%%%%%%%%

\end{document}